\documentclass[12pt]{article}
\usepackage{amsmath,amssymb,exscale,cancel}
\usepackage{slashed}
\allowdisplaybreaks
\usepackage{graphicx}
\usepackage{graphicx}
\usepackage{epsfig}
\usepackage{multicol}
\usepackage{color}
\usepackage{mathrsfs}
\usepackage{blindtext}
 \usepackage{fancyhdr}
\usepackage{hyperref}
\usepackage{cite}
\usepackage{mathtools}

\usepackage{floatrow}

\usepackage{graphicx}
\usepackage{sidecap}

\DeclareMathOperator*{\SumInt}{%
\mathchoice%
  {\ooalign{$\displaystyle\sum$\cr\hidewidth$\displaystyle\int$\hidewidth\cr}}
  {\ooalign{\raisebox{.14\height}{\scalebox{.7}{$\textstyle\sum$}}\cr\hidewidth$\textstyle\int$\hidewidth\cr}}
  {\ooalign{\raisebox{.2\height}{\scalebox{.6}{$\scriptstyle\sum$}}\cr$\scriptstyle\int$\cr}}
  {\ooalign{\raisebox{.2\height}{\scalebox{.6}{$\scriptstyle\sum$}}\cr$\scriptstyle\int$\cr}}
}

\usepackage[latin1]{inputenc} 

\usepackage{multicol}  
 
 \usepackage{titlesec}
 
 \usepackage{rotating,slashed,xcolor,amsfonts,expdlist,charter}

\numberwithin{equation}{section}

\usepackage{xcolor}
\usepackage{sectsty}

\usepackage{mdframed}
\usepackage{titletoc}

\numberwithin{equation}{section}

\usepackage{xcolor}
\usepackage{sectsty}

\usepackage{mdframed}
\usepackage{titletoc}

\definecolor{secnum}{RGB}{13,151,225}
\definecolor{ptcbackground}{RGB}{212,237,252}
\definecolor{ptctitle}{RGB}{0,177,235}

\titlecontents{lsection}
  [5.8em]{\sffamily}
  {\color{secnum}\contentslabel{2.3em}\normalcolor}{}
  {\titlerule*[1000pc]{.}\contentspage\\\hspace*{-5.8em}\vspace*{5pt}%
    \color{white}\rule{\dimexpr\textwidth-15.5pt\relax}{1pt}}

\usepackage{hyperref}
\hypersetup{colorlinks,bookmarksopen,bookmarksnumbered,citecolor=rossos,
linkcolor=redy,pdfstartview=FitH,urlcolor=green-go}
\usepackage{slashed}

\definecolor{blus}{cmyk}{1,0.9,0,0.1}
\definecolor{verdes}{cmyk}{0.99,0,0.59,0.65}
\definecolor{rossos}{cmyk}{0,1,1,0.55}
\definecolor{redy}{cmyk}{0,1,1,0.7}
\definecolor{greeny}{cmyk}{0.99,0,0.59,0.98}
\definecolor{green-go}{cmyk}{0.79,0,0.59,0.5}

\usepackage{titlesec}

\newcommand{\beq}{\begin{equation}}
\newcommand{\eeq}{\end{equation}}

\def\hhref#1{\href{http://arxiv.org/abs/#1}{arXiv:#1}} 

\newcommand{\tmtextbf}[1]{{\bfseries{#1}}}
\newcommand{\tmtextrm}[1]{{\rmfamily{#1}}}

\def\be{\begin{equation}}
\def\ee{\end{equation}}
\def\ba{\begin{array} }
\newcommand{\Tr}{\,{\rm Tr}}
\def\bac{\begin{array} {c}}
\def\bacc{\begin{array} {cc}}
\def\baccc{\begin{array} {ccc}}
\def\bacccc{\begin{array} {cccc}}
\def\ea{\end{array}}
\def\bea{\begin{eqnarray}}
\def\eea{\end{eqnarray}}

\definecolor{red}{rgb}{1,0,0}

\def\psl{\hbox{\hbox{${p}$}}\kern-1.9mm{\hbox{${/}$}}}
\def\dsl{\hbox{\hbox{${\partial}$}}\kern-2.2mm{\hbox{${/}$}}}
\def\Dsl{\hbox{\hbox{${D}$}}\kern-2.6mm{\hbox{${/}$}}}

\newcommand{\gappeq}{{\rlap{{\raise}.5ex\text{\ensuremath{>}}}{{\lower}.5ex\text{\ensuremath{\sim}}}}}
\newcommand{\lappeq}{{\rlap{{\raise}.5ex\text{\ensuremath{<}}}{{\lower}.5ex\text{\ensuremath{\sim}}}}}
\newcommand{\I}{\tmtextrm{1{\kern}-.24em l}}

\begin{document}
\topmargin -1.0cm
\oddsidemargin 0.9cm
\evensidemargin -0.5cm

{\vspace{-1cm}}
\begin{center}

\vspace{-1cm}

 {\tmtextbf{ 
 \hspace{-.cm}   
{\LARGE  \color{rossos}Fermion Thermal Field Theory for a Rotating Plasma}
\\
{\large (with Applications to Neutron Stars)}
 \hspace{-1.6cm}}} {\vspace{.5cm}}\\

\vspace{1.3cm}

{\large{\bf  Alberto Salvio }}

{\em  
\vspace{.4cm}
 Physics Department, University of Rome Tor Vergata, \\ 
via della Ricerca Scientifica, I-00133 Rome, Italy\\

\vspace{0.6cm}

I. N. F. N. -  Rome Tor Vergata,\\
via della Ricerca Scientifica, I-00133 Rome, Italy\\

  \vspace{0.5cm}

}
\vspace{1.5cm}
\end{center}

\noindent ---------------------------------------------------------------------------------------------------------------------------------
\begin{center}
{\bf \large Abstract}
\end{center}
\noindent   This paper provides a systematic and complete study of thermal field theory with fermion fields of any kind for generic equilibrium density matrices, which feature arbitrary values not only of temperature and chemical potentials, but also average angular momentum. This extends a previous study that focused on scalar fields, to all fermion-scalar theories. Both Dirac and Majorana fermions and both Dirac and Majorana masses are covered. A general technique to compute ensemble averages is provided. Path-integral methods are developed to study thermal Green's functions (with an arbitrary number of points) in generic interacting fermion-scalar theories, which cover both the real-time and imaginary-time formalism. These general results are applied to physical situations typical of neutron stars, which are often quickly rotating: the Fermi surface and Fermi momentum, the average energy, number
density and angular momentum for degenerate fermions and particle production (such as neutrino production from rotating neutron stars, e.g. pulsars). In particular, it is shown that the neutrino production rate due to the direct URCA (DU) processes grows indefinitely as the angular velocity approaches the inverse linear size of the plasma and, therefore, rotation can significantly increase this rate.

\vspace{0.7cm}

\noindent---------------------------------------------------------------------------------------------------------------------------------

\newpage

\tableofcontents

\noindent --------------------------------------------------------------------------------------------------------------------------------

\vspace{0.2cm}

\section{Introduction}\label{intro}

When applying the physical laws to cases of interest we often face difficulties due to  large numbers of particles. This can happen even when  relativistic  and/or quantum effects are important, like in compact astrophysical objects and/or in the early universe. In these situations one can combine relativity, quantum mechanics and statistics, to obtain thermal field theory (TFT). By now TFT is the standard theoretical tool to study particle physics processes (decays, scattering processes, particle production, phase transitions, etc.) in a medium (see~\cite{Bellac:2011kqa,Nair:2005iw} for textbooks,~\cite{Landsman:1986uw,Quiros:1994dr,Laine:2016hma} for monographs and~\cite{Salvio:2024upo} for an introduction from first principles).

At thermodynamic equilibrium, the density matrix, the key input in TFT, can be expressed in terms of all conserved quantities: the Hamiltonian, the linear and angular momentum and all conserved charges~\cite{LandauLifshitz}.  A previous paper~\cite{Salvio:2025rma} initiated, in the case of pure scalar theories, a systematic study of (generically interacting) TFT for the most general equilibrium density matrix, including not only temperature and chemical potentials associated with the conserved charges, but also a non-vanishing value of the average angular momentum.  The above-mentioned density matrix was carefully investigated by e.g.~\cite{Zubarev:1979afm,Weert,Becattini:2012tc} and important studies of some specific scalar TFT in the presence of a rotating plasma were performed by e.g.~Refs.~\cite{Vilenkin:1980zv,Buzzegoli:2017cqy,Prokhorov:2019yft,Becattini:2020qol,Kuboniwa:2025vpg,Siri:2025kdw}.

The literature also includes previous relevant studies of some specific fermion TFT in the presence of a rotating plasma~\cite{Vilenkin:1978hb,Vilenkin:1979ui,Vilenkin:1980zv,Ambrus:2015lfr,Chernodub:2016kxh,Buzzegoli:2017cqy,Prokhorov:2019yft,Buzzegoli:2020ycf,Palermo:2021hlf,Palermo:2023cup,Castano-Yepes:2025zae,Zhu:2025pxh}. However, a systematic study of (generically interacting) fermion TFT for the most general equilibrium density matrix is currently lacking.
Therefore, the purpose of the present  work is to fill this gap and extend the analysis of Ref.~\cite{Salvio:2025rma} to all kinds of fermions, including arbitrary values of  the average angular momentum, the temperature and all possible chemical potentials\footnote{See also  Ref.~\cite{Salvio:2026ewl} for a recent extension of Ref.~\cite{Salvio:2025rma} and the present work to general gauge theories.}. The extension to fermions is important because it can be applied, among other things, to neutron stars, which typically feature various types of fermions (neutrons, protons, electrons, etc.) and are often quickly rotating because of their small size (as theoretically anticipated in~\cite{Pacini:1967epn} and confirmed by the discovery of pulsars).  Then, another purpose of this work is  to apply the above-mentioned formalism to physical situations that are typically realized in neutron stars. 

However, for the sake of generality here both Dirac and Majorana fermions are studied and general mass terms, including Dirac and Majorana mass terms, are discussed. Majorana fermions with Majorana masses could be useful to study various extensions of the Standard Model such as those featuring a type-I see-saw. This type of sterile neutrinos, for example, could be emitted by rotating astrophysical compact objects made of several types of fermions.  

Clearly, the ensemble averages of observables are among the most important quantities that one can compute in TFT. However, very important are also the thermal Green's functions (the statistical average of the expectation values of the time-ordered product of a generic number of fields taken on a complete set of states): the applications of thermal Green's functions include, among other things, the determination of the effective action, which allows us, for example, to study possible phase transitions, and the computation of rates of particle processes (decays, scattering processes and particle production). Therefore, an important goal of this paper is to provide systematic techniques to determine the ensemble averages of observables and the thermal Green's function for the most general equilibrium density matrix in an arbitrary fermion-scalar TFT.

In the generically interacting case, the  path-integral approach can give us both these quantities. So  in this work the path integral representation of the partition function, which gives us the ensemble averages of observables, and of the Green's functions is investigated extending to general fermion-scalar theories the previous analysis of~\cite{Salvio:2025rma} valid for scalars only, both in the real- and imaginary-time formalism.    
Although this general formalism may hold at the non-perturbative level, here the tools to perform perturbation theory are provided too  (the propagators and how to combine them to form physical quantities in general  TFTs involving fermions for rotating plasmas with arbitrary equilibrium density matrices). 

Another purpose of this work is to furnish applications of those general results to several situations of physical relevance and featuring rotating plasmas with fermions. The relevant examples provided  here include the Fermi surface and Fermi momentum, the average energy, number density and  angular momentum for strongly degenerate fermions, particle production such as neutrino production from rotating neutron stars, etc. 

Moreover, several further motivations for these studies come to mind. Their applications can include phase transitions, decays, scattering processes and particle production around other compact objects, such as ordinary and primordial black holes and exotic compact objects. For instance, the accretion disks and coronas around black holes can be often considered rotating plasmas in approximate thermodynamic equilibrium.  Furthermore, one can conceive investigating the same phenomena (phase transitions, decays, scattering processes and particle production) in a lab, engineering a rotating plasma.

 The paper is organized as follows. 
\begin{itemize}
\item In the next section, as a first step towards the goals of this paper, the free field case is studied, keeping,
however, a general number of fermions (including both Dirac and Majorana fermions) and general values of particle masses (including both Dirac and Majorana masses), temperature,
chemical potentials and average angular momentum.
Keeping masses and chemical potentials general allows us to obtain formul\ae~that  are applicable to situations, typical of neutron stars (as discussed in Sec.~\ref{Direct URCA processes in rotating neutron stars}), where in-medium effects can be captured by effective masses and effective chemical potentials (leading to  
what one could call ``quasi-free fields").  The ensemble average of all relevant quantities and all 2-point functions  are investigated too. Special attention is devoted to cases of relevance for neutron stars. 
\item Sec.~\ref{Fermion path integral} is devoted to the derivation of the general path-integral formula for the partition function and the Green's functions, without committing ourselves to any specific underlying theory, but providing the most general expressions that are valid for any fermion-scalar theories. 
\item Finally, Sec.~\ref{Some applications} illustrates some applications of the general results previously obtained. In particular, that section discusses how the Fermi surface, the corresponding momenta and the weakly coupled fermion production rates are affected by rotation. Again, special attention is devoted to cases of relevance for neutron stars.  
\item Sec.~\ref{Conclusions} provides a detailed summary of the main original results of the paper and the final conclusions.
\end{itemize}

  \section{(Quasi-)free fields}\label{fermion Free fields}
  
  Let us start by considering a generic number of free Dirac  fields,  $\psi_s$, with Dirac masses. Later on also Weyl fields as well as Majorana masses will be studied. 
  The Lagrangian is here given by 
 \be \mathscr{L}= \bar \psi (i\slashed{\partial}-\mu_F) \psi,
  \label{freeLagf}\ee
where 
$\mu_F$ is the (Dirac) mass matrix related to the fermion squared mass matrix $M_F^2$ through $M_F^2\equiv \mu_F\mu_F^\dagger$. 
A vector notation is used, $\psi$ is an array of Dirac fields with components $\psi_s$ (where $s$ is a species index) and, as usual, $\bar\psi \equiv \psi^\dagger \gamma^0$ and $\slashed{\partial}\equiv \gamma^\mu \partial_\mu$, where  $\gamma^\mu$ are the Dirac matrices, which satisfy $\{\gamma^\mu,\gamma^\nu\}=2\eta^{\mu\nu}$. 
   
   An  internal (not necessarily Abelian) symmetry group $\mathcal{G}$ acts on $\psi$ as follows: 
\be \psi \to \exp(i\alpha_a t^a) \psi \label{GonF} \ee
for some real parameters $\alpha_a$, 
where the $t^a$ are the generators of $\mathcal{G}$  in the representation of fermions. The $t^a$ are Hermitian matrices and the invariance of the mass terms in~(\ref{freeLagf}) implies $[t^a,\mu_F]=0$, which in turn tells us (by Schur's Lemma) that $\mu_F$ can be taken to be block diagonal with each block proportional to the identity matrix;
 the different blocks correspond to irreducible representations of $\mathcal{G}$.  
 This allows us to consider, at least for free fields, the various irreducible representations separately as we do from now on in this Sec.~\ref{fermion Free fields}. All fields belonging to the same irreducible representation have of course the same mass, which in the following is denoted\footnote{The letter $m$ is not used for the mass here because it is used for the angular-momentum quantum number, see below.} $\mu$. 
 The generators of $\mathcal{G}$ in the given irreducible representation are denoted ${\cal R}^a$.
 
 The corresponding field operator $\Psi$ is the most general solution of the Dirac equation
\be (i\slashed{\partial}-\mu)\Psi = 0 \label{DEq}\ee 
satisfying the canonical anticommutation relations
\be \{\Psi_\alpha(t,\vec x), \Psi_\beta(t,\vec y)\} =0, \quad \{\Psi_\alpha(t,\vec x), \Psi_\beta^\dagger(t,\vec y)\} =\delta_{\alpha\beta} \delta(\vec x-\vec y). \label{CanAComm}\ee

 Now, choosing the reference frame appropriately (see Ref.~\cite{Salvio:2025rma} for all details), the most general equilibrium density matrix~\cite{LandauLifshitz,Zubarev:1979afm,Weert,Becattini:2012tc,Vilenkin:1980zv}, even at the fully interacting level, can always be written as follows:
 \be \rho = \frac{e^{-\beta (H-\vec\Omega  \cdot \vec J - \mu_a Q^a)}}{Z}, \label{rhoRest}
  \ee
  where $Z$ is the partition function, $\beta\equiv1/T$ (the inverse of the temperature), $H$ is the Hamiltonian, $\vec J$ is the angular momentum, the $Q^a$ are the full set of charges, which generate the internal  symmetry group $\mathcal{G}$, and $\mu_a$ is the chemical potential associated with $Q^a$. Also, $\vec\Omega$ is another thermodynamical quantity associated with the average angular momentum of the system. Sometimes $\vec\tau \equiv -\beta \vec\Omega$  is named thermal vorticity.   As shown in~\cite{Salvio:2025rma} for scalars and later on in Sec.~\ref{Fermion path integral} for fermions,  $\vec\Omega$ can be identified with the angular-velocity vector of  a rigidly-rotating plasma. This was also previously noted by Refs.~\cite{Vilenkin:1978hb,Vilenkin:1979ui,Vilenkin:1980zv,Buzzegoli:2017cqy,Becattini:2020qol}.
  
Let us take now the cylindrical coordinates
\be x^1 = r\cos\phi, \qquad x^2 = r\sin\phi, \qquad x^3 = z,\label{CyCoo} \ee
with the third axis identified with the rotation axis. 
 One can work in the basis of eigenstates of the commuting operators $H$, $P^z$, $J_z$ and $\vec J\cdot \vec P/|\vec p|$, where the $P^i$ and $J_i$ are the components of the linear and angular momentum, $\vec J\cdot \vec P/|\vec p|\equiv J_iP^i/|\vec p|$  is the helicity and $|\vec p|$ is the length of the linear three-momentum. Let  us call $q$ the corresponding set of eigenvalues, which for the  fermion in question are  $\omega$, $p$, $m+1/2$ (with $m$ being  a generic integer) and $\sigma =\pm 1/2$, respectively.  Note that when $\mu=0$ one can consider just spinors that are eigenstates of chirality (twice the helicity, i.e.~$\gamma_5$ in the choice of~\cite{BjorkenDrell}) with a definite eigenvalue.
In general, one can write 
 \be \Psi_s(x) = \SumInt_q (\mathcal{U}_q(x) c_{qs}+ \mathcal{V}_q(x) d^\dagger_{qs}), \label{FreeFexp}\ee 
 where the integro-sum over $q$ is now defined, for any integrand $[...]$, by
\be \SumInt_q [...]\equiv \sum_{m=-\infty}^{+\infty}\sum_{\sigma=\pm1/2}\int_\mu^\infty d\omega\int_{-p_0}^{p_0}dp \left[...\right],
 \ee 
 $p_0\equiv \sqrt{\omega^2 -\mu^2}$, 
 the $\mathcal{U}_q$ and $\mathcal{V}_q$ are the complete set of solutions of~(\ref{DEq}) in this basis for particles and antiparticles, respectively,  
 and  the $c_{qs}$ and $d_{qs}$ are the corresponding annihilation operators for the fermion and antifermion, respectively, of species $s$. They satisfy
  \bea &&\{c_{q s},c_{q' s'}^\dagger\} = \{d_{q s},d_{q' s'}^\dagger\} =\delta(q-q')\delta_{ss'}, \label{antNot}\\ 
&&\{c_{q s},c_{q' s'}\}=\{d_{q s},d_{q' s'}\}=\{c_{q s},d_{q' s'}\}=\{c_{q s},d_{q' s'}^\dagger\}=0,\label{ant} \eea 
 where 
 \be \delta(q-q') \equiv  \delta_{mm'}\delta_{\sigma\sigma'}\delta(\omega - \omega')\delta(p-p'). \ee 
 Since the $\mathcal{U}_q$ and $\mathcal{V}_q$ form a  complete set of eigenfunctions of Hermitian operators (corresponding to the Hamiltonian, the linear and angular momentum along the third axis and the helicity)
  they can be normalized in a way that 
 \be \int d^3 x \, \mathcal{U}^\dagger_{q'}(x)\mathcal{U}_{q}(x) = \delta(q'-q)  = \int d^3 x \, \mathcal{V}^\dagger_{q'}(x)\mathcal{V}_{q}(x), \qquad  \int d^3 x \, \mathcal{V}^\dagger_{q'}(x)\mathcal{U}_{q}(x) =0  \label{OrtoF}\ee 
 and also 
 \be \SumInt_q( \mathcal{U}_q(t,\vec x)\mathcal{U}_q^\dagger(t,\vec y) +  \mathcal{V}_q(t,\vec x)\mathcal{V}_q^\dagger(t,\vec y)) = \delta(\vec x-\vec y). \label{completeF}\ee 
Here $^\dagger$ represents the conjugate transpose, so the quantity in~(\ref{completeF}) is a matrix in the spinor space. Using~(\ref{antNot}),~(\ref{ant}) and~(\ref{completeF}), one can check the anticommutation relations in~(\ref{CanAComm}). 
 The $\mathcal{U}_q$ and the $\mathcal{V}_q$ contain cylindrical Bessel functions. A way to see this is to note that $\Psi$, just like the free scalar operator $\Phi$, satisfies the Klein-Gordon equation.

 The explicit form of the $\mathcal{U}_q$ and $\mathcal{V}_q$ depend on the basis choice for the $\gamma^\mu$. The explicit expression of $\mathcal{U}_q$ and $\mathcal{V}_q$ in the choice, for example, of~\cite{BjorkenDrell} was obtained in\footnote{Here a different normalization is used to implement~(\ref{OrtoF}) and~(\ref{completeF}), so the solutions denoted $U_j$ and $V_j$ for Dirac fermions  in~\cite{Ambrus:2015lfr} are related to $\mathcal{U}_q$ and $\mathcal{V}_q$ through $\mathcal{U}_q =\sqrt{\omega} U_j$ and $\mathcal{V}_q=\sqrt{\omega} V_j$.}~\cite{Ambrus:2015lfr} (see also the previous study~\cite{Vilenkin:1979ui} for the case of chiral spinors). For this choice $\mathcal{V}_q = i \gamma^2 \mathcal{U}_q^*$.

 So far, only Dirac masses have been considered. In the limit where some of these masses vanish one can consider chiral fermions with just one helicity state (either $\sigma=+1/2$ or $\sigma=-1/2$) as a particular case. In order to present an analysis that is as general as possible, let us now also include  Majorana masses in the analysis. These appear in well-motivated extensions of the SM, such as those featuring the type-I see-saw mechanism. The most suitable formalism to include Majorana masses is that of Weyl spinors,  which feature the following Lagrangian density 
 \be \mathscr{L}= \bar
 \psi i \slashed{\partial}\psi+\frac12 \left(\psi \mu_F\psi +\mbox{h.c.}\right).
  \label{freeLagW}\ee
 In the Weyl formalism we adopt the following notation.
\begin{itemize}
\item $\psi$ and $\bar\psi$ are two-component spinors with components $\psi_\alpha$ and $\bar\psi^\alpha$, respectively, ($\bar\psi^\alpha$ is interpreted here as the Hermitian conjugate of $\psi_\alpha$)
and the transpose operation is understood. We also introduce 
$\psi^\alpha\equiv \psi_\beta \epsilon^{\beta\alpha}$, $\bar\psi_\alpha\equiv \epsilon_{\alpha\beta}\bar\psi^\beta$, 
where $\epsilon^{\alpha\beta}$ and $\epsilon_{\alpha\beta}$ are the antisymmetric symbols with $\epsilon^{12}=1$ and $\epsilon_{12}=-1$, such that $\epsilon^{\alpha\beta}\epsilon_{\beta\gamma} = \delta^\alpha_{~\gamma}$.
\item The kinetic term $\bar
 \psi i \slashed{\partial}\psi$ is now constructed with the $2\times 2$ matrices $\bar\sigma^\mu$ (defined by $\{\bar{\sigma}^{\mu}\}\equiv (1, -\vec{\sigma})$ and $\vec{\sigma}$ represents the three Pauli matrices) as 
 \be\bar
 \psi i \slashed{\partial}\psi \equiv \bar
 \psi i \bar\sigma^\mu \partial_\mu\psi \equiv\bar\psi^\alpha i\bar\sigma^{\mu~\beta}_{~\alpha} \partial_\mu\psi_\beta. \ee
 \item Finally, $\mu_F$
 is the fermion mass matrix, which can include both Dirac and Majorana masses, and 
 \be \psi \mu_F \psi \equiv  \psi_{\beta i}\epsilon^{\beta\alpha} \mu_F^{ij} \psi_{\alpha j}, \qquad (\psi \mu_F\psi)^\dagger = \bar\psi_{\alpha j} (\mu_F^{ij})^* \epsilon^{\beta\alpha}\bar\psi_{\beta i}\equiv \bar\psi \mu_F^\dagger \bar\psi. \label{GenMass} \ee
 \end{itemize}
The field $\psi$ is assumed to transform under $\mathcal{G}$ as in~(\ref{GonF}), in a way that also the mass terms above are invariant. Note that we can put  $\mu_F$ in diagonal form through a unitary transformation\footnote{This is known as the complex Autonne-Takagi factorization, see e.g.~\cite{Youla}.} acting on $\psi$.  
 We then  work with a field basis where $\mu_F$ is diagonal. The absolute values of the diagonal elements of $\mu_F$ are the fermion  masses. 
 Also in this case, in a given  irreducible representation of $\mathcal{G}$ all particles have the same masses by Schur's Lemma because the Hamiltonian always commutes with the $Q^a$ (the $Q^a$ are assumed to be conserved).
  The mass and generators in the given irreducible representation are denoted again $\mu$ and ${\cal R}^a$, respectively.  From now on in this Sec.~\ref{fermion Free fields} the various irreducible representations  are considered separately also for Weyl fields. This formalism allows us to easily describe Majorana fermions too.
 
The field operator $\Psi$ for a massive\footnote{The massless case can be easily treated through Dirac fields by considering eigenspinors  of chirality.} Majorana fermion is a solution of
\be i \slashed{\partial}\Psi = -\mu \epsilon \bar\Psi \label{WeylEq}\ee
instead of~(\ref{DEq}), satisfying the  canonical anticommutation relations
\be \{\Psi_\alpha(t,\vec x), \Psi_\beta(t,\vec y)\} =0, \quad \{\Psi_\alpha(t,\vec x), \bar \Psi^\beta(t,\vec y)\} =\delta_\alpha^{~\beta} \delta(\vec x-\vec y). \label{CanACommW}\ee
 Here $\epsilon$ is the $2\times2$ antisymmetric matrix with $\epsilon_{12}=-1$. When working with Weyl-spinor operators, $\bar\Psi^\alpha$ represents the Hermitian conjugate of $\Psi_\alpha$.  In the case of Majorana fermions, which are described here by Weyl spinors for convenience, the decomposition of a general $\Psi_s$ (with $s$ being the species index) in terms of annihilation and creation operators of particle states with definite values of  $H$, $P^z$, $J_z$ and $\vec J\cdot \vec P/|\vec p|$ reads
    \be \Psi_s(x) = \SumInt_q (X_q(x) a_{qs}+Y_q(x) a^\dagger_{qs}), \label{FreeWexp}\ee
    where the $X_q$ and $Y_q$ are eigenspinors of  $H=i\partial_t$, $P^z$, $J_z$ and $\vec J\cdot \vec P/|\vec p|$ with eigenvalues $\{\omega,p,m+1/2,\sigma\}$ and $\{-\omega,-p,-m-1/2,\sigma\}$, respectively.  
    Eq.~(\ref{WeylEq}) then implies
    \be (\omega+2|\vec p|\sigma) X_q +\mu \epsilon \bar Y_q =  (-\omega+2|\vec p|\sigma) Y_q +\mu \epsilon \bar X_q = 0, \label{EOMXY}\ee 
    where a bar on top of bispinors  represents a complex conjugate.
    Combining these two equations and using $\sigma=\pm 1/2$ leads to the on-shell relation $\omega^2=\mu^2+\vec p^2$. Moreover, both these equations allow us to express $Y_q$ in terms of $X_q$ (and viceversa):
    \be Y_q = \frac{\omega+2|\vec p|\sigma}{\mu} \epsilon \bar X_q. \label{YfromX}\ee
     
     Note that the Weyl field operator $\Psi_s$ in~(\ref{FreeWexp}) features only one type of annihilation operators, $a_{qs}$,
     as appropriate for Majorana fermions. Being a fermion system
 \be    \{a_{q s},a_{q' s'}^\dagger\} = \delta(q-q')\delta_{ss'}, \qquad \{a_{q s},a_{q' s'}\}  = 0 . \label{antNotW} \ee

 One can easily show that the Majorana field $\Psi$ in~(\ref{FreeWexp}) satisfies the Klein-Gordon equation $\partial^2\Psi=-\mu^2\Psi$. As a result both $X_q$ and $Y_q$ are eigenfunctions of $\vec P^2$. Since the $X_q$  form a  complete set of eigenfunctions of Hermitian operators (corresponding to $\vec P^2$, 
$P^z$, $J_z$ and $\vec J\cdot \vec P/|\vec p|$)
  they can be normalized in a way that 
 \be \int d^3 x \, \bar X_{q'}(x)X_{q}(x) = \frac{\mu c_x}{2\omega} \delta(q'-q),  \quad \mbox{with} \quad c_x \equiv \sqrt{\frac{\omega-2|\vec p|\sigma}{\omega+2|\vec p|\sigma}},  \label{OrtoFX}\ee 
 which corresponds to the completeness relation
 \be \SumInt_q X_{q\alpha}(t,\vec x)\bar X_q^\beta(t,\vec y) = \frac{\mu c_x}{2\omega}\delta_\alpha^{~\beta} \delta(\vec x-\vec y). \label{completeX}\ee 
 Using Eqs.~(\ref{EOMXY}) one finds the corresponding relations for the $Y_q$:
 \be \int d^3 x \, \bar Y_{q'}(x)Y_{q}(x) = \frac{\mu}{2\omega  c_x} \delta(q'-q), \label{OrtoFY}\ee
 and
 \be \SumInt_q Y_{q\alpha}(t,\vec x)\bar Y_q^{\beta}(t,\vec y) = \frac{\mu}{2\omega  c_x}\delta_\alpha^{~\beta} \delta(\vec x-\vec y). \ee 
 Using~(\ref{YfromX}),~(\ref{antNotW}) and~(\ref{completeX}) one can check the  canonical anticommutation relations in~(\ref{CanACommW}).
 With the normalization used in~(\ref{OrtoFX}), we found that $2\pi \exp(i\omega t-i p z)X_q$ can be taken to be proportional to the quantitiy $\phi_j$ computed in~\cite{Ambrus:2015lfr}  with the proportionality factor given by $\sqrt{\mu c_x/2}$. Having determined $X_q$, the other  eigenspinor $Y_q$ is given by~(\ref{YfromX}).

   \subsection{Computing ensemble averages} \label{Computing ensemble averages f}

In this section we provide a general method to  compute averages in the case of free fermion fields, i.e.~for systems involving fermions with negligibly small interactions.  
However, as discussed in Sec.~\ref{Direct URCA processes in rotating neutron stars}, in some cases one can take into account in-medium effects  by substituting  the masses with effective masses and the chemical potentials with effective chemical potentials (leading to what one could call ``quasi-free fields"). Let us suppose that this substitution is performed in this Sec.~\ref{Computing ensemble averages f}. Sec.~\ref{Fermion path integral} will then furnish the methods to address theories with general interactions.

To facilitate the computation of averages let us perform a change of basis in the space of particle states. This can  be done as follows. 
Note that $\mathcal{G}$ acts on $\Psi(x)$ as
\be \exp(i\alpha_aQ^a)\Psi(x)\exp(-i\alpha_aQ^a) = \exp(i\alpha_a{\cal R}^a)\Psi(x). \label{GonPsi} \ee
Let us first consider Dirac fermions, 
Eq.~(\ref{FreeFexp}). We will later clarify what has to be modified in the case of Majorana fermions, Eq.~(\ref{FreeWexp}).
Note that (\ref{GonPsi}) corresponds to the following action on the annihilation and creation operators,
\bea  \exp(i\alpha_aQ^a)c_{qs}\exp(-i\alpha_aQ^a) = \exp(i\alpha_a{\cal R}^a)_{ss'}c_{q s'},\label{Gonacf} \\ 
\exp(i\alpha_aQ^a)d_{q s}\exp(-i\alpha_aQ^a) = \exp(i\alpha_a\bar {\cal R}^a)_{ss'}d_{q s'}, \label{Gonadf} \eea
where $\bar {\cal R}^a  = - ({\cal R}^a)^*$. 
The transformation rules in~(\ref{Gonacf}) and~(\ref{Gonadf}) imply the following action of $\mathcal{G}$  on one-particle states, $|q,s\rangle\equiv c_{q s}^\dagger|0\rangle$ and $|\overline{q,s}\rangle\equiv d_{q s}^\dagger|0\rangle$  (these states have energy $\omega$, linear and angular  momentum along the third axis $p$ and $m+1/2$, respectively, helicity $\sigma$ and species $s$):
  \be  \exp(i\alpha_aQ^a) |q,s\rangle =  \exp(i\alpha_a\bar {\cal R}^a)_{ss'} |q,s'\rangle, \quad \exp(i\alpha_aQ^a) |\overline{q,s}\rangle =  \exp(i\alpha_a{\cal R}^a)_{ss'} |\overline{q,s'}\rangle,\label{Gon1Pf} \ee
  where  the invariance of the vacuum $|0\rangle$ under $\mathcal{G}$ was used.
The expressions above imply, among other things, 
  \be \mu_a Q^a|q,s\rangle =  (\mu_a \bar {\cal R}^a)_{ss'} |q,s'\rangle, \qquad \mu_a Q^a|\overline{q,s}\rangle =  (\mu_a {\cal R}^a)_{ss'} |\overline{q,s'}\rangle. \ee
    Now, by performing a $\mu_a$-dependent unitary transformation of these states,
  \be |q;d\rangle\equiv W_{ds} |q,s\rangle, \quad |\overline{q;d}\rangle\equiv W^*_{ds} |\overline{q,s}\rangle \ee 
  (with the $W_{ds}$ satisfying $W_{ds}W^*_{d's}= \delta_{dd'}$) it is possible to diagonalize both $\mu_a\bar {\cal R}^a$ and $\mu_a{\cal R}^a$:
  \be W\mu_a \bar {\cal R}^aW^\dagger = \mathcal{M}^F, \quad W^*\mu_a {\cal R}^aW^T = -\mathcal{M}^F, \label{fromttoD} \ee 
  where $W$ is the matrix with elements $W_{ds}$ and $\mathcal{M}^F$ is a (generically $\mu_a$-dependent) diagonal real matrix. Therefore, in the new   basis 
    \be \mu_a Q^a|q;d\rangle =  \mathcal{M}^F_d |q; d\rangle, \quad \mu_a Q^a|\overline{q;d}\rangle =  -\mathcal{M}^F_d |\overline{q; d}\rangle, \label{muQMf}\ee
    where the $\mathcal{M}^F_d$ are the diagonal elements of $\mathcal{M}^F$ and in the right-hand sides of~(\ref{muQMf}) there is no sum over the index $d$. The $\mathcal{M}^F_d$ encode the effect of the chemical potentials for an arbitrary (Abelian or non-Abelian) symmetry group $\mathcal{G}$.

    Following~\cite{Salvio:2025rma}, let us discretize the variables  $\omega$ and $p$  such that integrals over these quantities become sums, for example, by putting the system in a cylinder of height $L$ and radius $R$; this effectively divides the ranges of $\omega$ and $p$ in small discrete steps of size $\Delta\omega$ and  $\Delta p$. One can then let $\Delta\omega\to0$ and $\Delta p\to0$ to recover the continuum case. Moreover, one can introduce the rescaled annihilation operators $\gamma_{q, d} \equiv \sqrt{\Delta\omega\Delta p} \, c_{q, d}$ and $\delta_{q, d} \equiv \sqrt{\Delta\omega\Delta p} \, d_{q, d}$ (with $c_{q,d}\equiv W_{ds}^*c_{q s}$ and $d_{q,d}\equiv W_{ds}d_{q s}$). The only non-vanishing anticommutators between rescaled annihilation and creation operators are
    \be \{\gamma_{q, d}, \gamma_{q', d'}^\dagger\} = \delta_{qq'}\delta_{dd'}, \quad \{\delta_{q, d}, \delta_{q', d'}^\dagger\} = \delta_{qq'}\delta_{dd'}, \label{ACommccdD}\ee
    with $\delta_{qq'}\equiv \delta_{mm'}\delta_{\sigma\sigma'} \delta_{\omega \omega'}\delta_{pp'}$. This discretization is useful to easily compute $\rho$, $Z$ and the ensemble average of relevant quantities in full generality. The continuum limit (which corresponds to the large-volume limit in coordinate space) will be taken afterwards.

  The density matrix in~(\ref{rhoRest}) in the fermion case  can then be expressed in terms of the number operators for particles and antiparticles, respectively 
    \be N_{q d} \equiv \gamma^\dagger_{q, d}\gamma_{q, d}, \quad \bar N_{q d} \equiv \delta^\dagger_{q, d}\delta_{q, d}\ee 
    as follows:
    \be \rho=\frac1{Z} \exp\left(-\beta \sum_{q d}\left\{\left[ \omega  -(m+1/2)\Omega   - \mathcal{M}_d^F  \right]N_{q d}+\left[ \omega  -(m+1/2)\Omega   + \mathcal{M}_d^F  \right]\bar N_{q d}\right\} \right). \label{rhofref} \ee 
    Recall that the  $\mathcal{M}_d^F$  represent the contribution of a general set of chemical potentials $\mu_a$ in the fermion case. The quantity in~(\ref{rhofref}) is nothing but the density matrix with zero thermal vorticity and chemical potentials, but with energies $\omega$ replaced by $\omega - (m+1/2)\Omega - \mathcal{M}_d^F$ for particles and by $\omega - (m+1/2)\Omega + \mathcal{M}_d^F$ for antiparticles. As a result the partition function is 
    \be Z =  \left[\prod_{q d} \left(1+e^{-\beta(\omega - (m+1/2)\Omega - \mathcal{M}_d^F)}\right)\right] \left[\prod_{q d} \left(1+e^{-\beta(\omega - (m+1/2)\Omega + \mathcal{M}_d^F)}\right)\right], \label{Zpartf}\ee
    where the first square bracket refers to particles and the second one to antiparticles.
    Then, using  
\be \log Z =   \sum_{q d} \left(\log \left(1+e^{-\beta(\omega - (m+1/2)\Omega - \mathcal{M}_d^F)}\right)
+\log \left(1+e^{-\beta(\omega - (m+1/2)\Omega + \mathcal{M}_d^F)}  \right)\right), \label{logZf}\ee 
one finds that the only non-vanishing averages of products of two annihilation and creation operators are\footnote{For a generic operator $\mathcal{F}$ the ensemble average is 
$\langle \mathcal{F}\rangle =  \Tr(\rho \mathcal{F})$.
} 
\bea \langle \gamma^\dagger_{q, d}\gamma_{q', d'}\rangle &=& f_F(\omega - (m+1/2)\Omega - \mathcal{M}_d^F)\delta_{dd'} \delta_{qq'}, \label{dgamma}\\ 
\langle \delta^\dagger_{q, d}\delta_{q', d'}\rangle &=& f_F(\omega - (m+1/2)\Omega + \mathcal{M}_d^F) \delta_{dd'} \delta_{qq'}, \label{ddelta}
\\ 
\langle \gamma_{q, d}\gamma^\dagger_{q', d'}\rangle &=& (1-f_F(\omega - (m+1/2)\Omega - \mathcal{M}_d^F)) \delta_{dd'} \delta_{qq'}, \label{gammad}\\ \langle \delta_{q, d}\delta^\dagger_{q', d'}\rangle &=& (1-f_F(\omega - (m+1/2)\Omega + \mathcal{M}_d^F)) \delta_{dd'} \delta_{qq'}, \label{deltad}\eea
where
\be f_F(x) \equiv \frac1{e^{\beta x}+1} \label{fFdef}\ee
is the Fermi-Dirac distribution. Setting $d=d'$ and $q=q'$ in~(\ref{dgamma}) and~(\ref{ddelta}) one finds the average numbers of fermions and antifermions, recovering the results of~\cite{Vilenkin:1978hb} for vanishing chemical potentials.

In the small-temperature limit, which is  typically relevant, for example, for neutron stars, $f_F(x)\simeq \theta(-x)$, where $\theta(x)$ is the Heaviside step function. Therefore, the average numbers go to 1 or 0 in this limit for $\omega - (m+1/2)\Omega  <\pm\mathcal{M}_d^F$ or $>\pm\mathcal{M}_d^F$, respectively, where the plus and minus signs refer to fermions and antifermions, respectively. This is the case when the (anti)fermions are strongly degenerate. The presence of $\Omega$ in these inequalities leads to a deformation of the Fermi surface that one has in a non-rotating fermion plasma. Such deformation will be discussed in Sec.~\ref{Fermi momentum and Fermi surface}.

Let us recall that in~(\ref{Zpartf}) a single fermion irreducible representation of $\mathcal{G}$ is considered: to obtain the partition function for all  irreducible representations one can simply take the product of all partition functions of single irreducible representations.

Moreover, using~(\ref{logZf}) and the general expressions of $\langle H\rangle$,~$\langle  J_i\rangle$ and~$\langle Q^a\rangle$ in~\cite{Salvio:2025rma}, the average values of $H$, $\vec J$  and $Q^a$ turn out to be 
\bea \langle H\rangle &=&  
\sum_{q d} \omega \left(f_F(\omega - (m+1/2)\Omega - \mathcal{M}_d^F)+f_F(\omega - (m+1/2)\Omega + \mathcal{M}_d^F)\right),\label{Hf} \\   \langle  J_z\rangle &=&  
\sum_{q d} \left(m+\frac12\right) \left(f_F(\omega - (m+1/2)\Omega - \mathcal{M}_d^F)+f_F(\omega -(m+1/2)\Omega + \mathcal{M}_d^F)\right), \label{Jzf}\\
\langle Q^a\rangle&=& 
\sum_{q d} \frac{\partial\mathcal{M}_d^F}{\partial\mu_a}\left(  f_F(\omega -(m+1/2)\Omega - \mathcal{M}_d^F)-  f_F(\omega -(m+1/2)\Omega + \mathcal{M}_d^F)\right) \label{avQf}.\eea
In the limit $\Delta p\to0$ (which corresponds to $L\to\infty$) these expressions allow us to compute  the average values of the energy, angular momentum and charges per unit of length in the $z$ direction in terms of integrals rather than sums over $p$. 

We can also go back to the original basis by inverting~(\ref{fromttoD}) and obtain, starting from~(\ref{dgamma})-(\ref{deltad}), that the only non-vanishing averages of pairs of annihilation and creation operators are\footnote{Functions of the matrices $\omega - (m+1/2)\Omega - \mu_a \bar {\cal R}^a$ and $\omega - (m+1/2)\Omega - \mu_a  {\cal R}^a$  can be computed with the spectral decomposition of those matrices as illustrated in~\cite{Salvio:2025rma}.}
\bea \langle \gamma^\dagger_{qs}\gamma_{q's'}\rangle &=& f_F(\omega - (m+1/2)\Omega - \mu_a \bar {\cal R}^a)_{ss'} \delta_{qq'}, \label{dgammad}\\ 
\langle \delta^\dagger_{qs}\delta_{q's'}\rangle &=& f_F(\omega - (m+1/2)\Omega - \mu_a  {\cal R}^a)_{ss'} \delta_{qq'}, \label{ddeltad}
\\ 
\langle \gamma_{qs}\gamma^\dagger_{q's'}\rangle &=& (1-f_F(\omega - (m+1/2)\Omega - \mu_a \bar {\cal R}^a))_{s's} \delta_{qq'}, \label{gammadd}\\ \langle \delta_{qs}\delta^\dagger_{q's'}\rangle &=& (1-f_F(\omega - (m+1/2)\Omega - \mu_a {\cal R}^a))_{s's} \delta_{qq'}, \label{deltadd}\eea
with $\gamma_{q s} \equiv W_{ds}\gamma_{q, d}$, $\delta_{q s} \equiv W^*_{ds}\delta_{q, d}$.
Corresponding averages for density matrices that also include, unlike~(\ref{rhoRest}), the linear momentum and the generators of Lorentz boosts were presented in~\cite{Palermo:2021hlf}.
Here, a closed form is found for density matrices of the form~(\ref{rhoRest}) in the presence of an arbitrary number of chemical potentials. Note that the expressions in~(\ref{dgammad})-(\ref{deltadd}) hold both for Abelian and non-Abelian internal symmetry groups.

Taking into account the expression~(\ref{logZf}), one finds that the convergence of the fermion averages, just like the convergence of the  averages for scalars~\cite{Salvio:2025rma}, requires the bound $\Omega<1/R$, so that 
\be v\equiv \Omega R\in[0,1), \label{vbounds}\ee
which agrees with the fact that the particles in the rotating plasma must not exceed the speed of light. For this reason the upper bound in~(\ref{vbounds}) can be interpreted as a causality bound.
In order to respect this bound here the large-$R$ limit is taken together with the small-$\Omega$ limit, keeping $v$ fixed in the interval $[0,1)$. Adapting the corresponding discussion in~\cite{Salvio:2025rma} for scalars, in the fermion case one  finds the following general formul\ae~for the averages of  the energy density $\rho_E$,  the angular momentum density per  unit of distance from the
 rotation axis, $\mathcal{J}_z$, and the charge densities $\rho_a$, which are  defined as the large-$R$ and large-$L$ limit of  $\langle H\rangle/(\pi R^2 L)$, $\langle  J_z\rangle/(\pi R^3 L)$ and $\langle Q^a\rangle/(\pi R^2 L)$,  respectively:
 \bea \langle \rho_E\rangle &=& 2\sum_d\int \frac{\alpha\zeta(\xi) d\alpha d\xi dp}{2\pi^2} \,  \omega \left(f_F (\omega -  v \alpha \xi- \mathcal{M}_d^F)+f_F (\omega -  v \alpha \xi+ \mathcal{M}_d^F)\right),  \label{rhoEpf}\\
 \langle \mathcal{J}_z\rangle &=&   2\sum_d\int \frac{\alpha \zeta(\xi)d\alpha d\xi dp}{2\pi^2} \,  \alpha \xi \left(f_F (\omega -  v \alpha \xi- \mathcal{M}_d^F)+f_F (\omega -  v \alpha \xi+ \mathcal{M}_d^F)\right), \label{CallJzf} \\ 
 \langle\rho_a\rangle &=&   2\sum_d\frac{\partial\mathcal{M}^F_d}{\partial\mu_a}\int \frac{\alpha\zeta(\xi)d\alpha d\xi dp}{2\pi^2} \, \left(f_F (\omega -  v \alpha \xi- \mathcal{M}_d^F)-f_F (\omega -  v \alpha \xi+ \mathcal{M}_d^F)\right),\label{rhoaf} \eea
 where  $\omega = \sqrt{\mu^2+\alpha^2 +p^2}$ and  the integral is over the full momentum space, $\alpha\in[0,\infty), \xi\in[-1,1], p\in (-\infty,\infty)$. The function $\zeta$ is given in~\cite{Salvio:2025rma,dataset}. Also, in~(\ref{rhoaf}) the overall factors of 2 are due to the sums over the two helicity states. In the case of massless fermions that have just one helicity state, those factors of 2 are thus absent. Some explanations are in order to illustrate how to obtain~(\ref{rhoEpf})-(\ref{rhoaf}). As illustrated, for example, in~\cite{Ambrus:2015lfr}, the required boundary conditions for fermions do not lead exactly to the same discrete values $\alpha_{m,n}$ of $\alpha=\sqrt{\omega^2-p^2-\mu^2}$ as in the scalar case for finite values of $m$. However,  to obtain~(\ref{rhoEpf})-(\ref{rhoaf}) one only needs the large-$m$ limit: this is because~(\ref{rhoEpf})-(\ref{rhoaf}) are obtained from~(\ref{Hf})-(\ref{avQf}) taking the large-$R$ limit and dividing by the volume $\pi R^2 L$ (and by an extra factor of $R$ in the case of $\langle \mathcal{J}_z\rangle$ in~(\ref{CallJzf})). Indeed, one can absorb some factors of $1/R$ in the definition of a continuous integration variable $y\equiv m/R$, which  takes finite values in the large-$R$ limit only for large $m$. On the other hand, when $m$ is large, the discrete values $\alpha_{m,n}$ for fermions do approach those for scalars
\be \alpha_{m,n} \to \frac{j_{m,n}}{R}, \label{alphadef}\ee
   where $j_{m,n}$ is the $n$th positive zero of the cylindrical Bessel function $J_m$.  A way to see this explicitly is to note that, although some components of the spinors in Eq.~(2.15) of~\cite{Ambrus:2015lfr} have values of $m$ that differ by one units, this difference does not matter in the large-$R$ limit as $(m+1)=(y+1/R)R$ (which approaches $yR$ in the large-$R$ limit).  For  large $m$ one has $\alpha_{m,n}\geq \alpha_{m,1}= |m|/R$, where the modulus appears because $j_{m,n}=j_{-m,n}$. So  we can substitute $m/R\to y$, where $y\in[-\alpha,\alpha]$. The $y$-independent  variation of $\alpha$  is then 
\be \Delta\alpha_{m,n}\equiv \frac{j_{m,n+1}-j_{m,n}}{R}. \label{Deltaalpha}\ee
 For large $n$ and fixed $m$ (which corresponds to $|y|\ll\alpha$) the variable $\alpha$ (see~(\ref{alphadef})) acquires non-vanishing values even in the large-$R$ limit  and, using McMahon's  Asymptotic Expansions of the $j_{m,n}$ for large $n$, one finds $R\Delta\alpha_{m,n}\to \pi$. 
 On the other hand, using asymptotic expansions of the $j_{m,n}$ for large $m$ and fixed $n$ (which corresponds to $|y|\simeq\alpha$) one finds that  $R\Delta\alpha_{m,n}$ grows indefinitely in this limit and so its inverse tends to zero.
  For arbitrary values of $y$, $R\Delta\alpha_{m,n}$ goes to an even function of the dimensionless ratio $y/\alpha$, which is what we call $\zeta^{-1}(y/\alpha)$. Finally, defining the variable $\xi\equiv y/\alpha$ one obtains the integration measure in~(\ref{rhoEpf})-(\ref{rhoaf}) and the integration domain specified just below~(\ref{rhoEpf})-(\ref{rhoaf}).
 
Physically, the transition from the discrete variable $m$ to the continuous variable $y$ should be understood as a convenient approximation to describe macroscopic or even astrophysical objects starting from microscopic laws. The results obtained in~(\ref{rhoEpf})-(\ref{rhoaf}) remain valid for rotating systems even if the large-$R$ limit corresponds to the small-$\Omega$ limit: in this approximation one trades the angular velocity $\Omega$ with the velocity at the radial boundary, $v$. But for rigidly-rotating plasmas, which is what we are considering here, a non-vanishing velocity at the radial boundary implies a non-vanishing velocity at any finite distance from the rotation axis, although such velocity tends to zero on the rotation axis. One can also note that approximating $m/R$ with a continuous variable corresponds to a sort of semi-classical approximation: in the exponent of the density matrix the angular-momentum quantum number $m$  appears multiplied by $\beta \Omega \hbar$ (having restored $\hbar$). So substituting $m/R$ with a continuous variable is valid when $\hbar \ll 1/(\beta\Omega)$.
 
Despite the Tolman law~\cite{Tolman:1930zza}, which implies that in a gravitational field (or in a non-inertial frame) an appropriately defined temperature $T_0$ is not homogeneous (see also Ref.~\cite{Becattini:2012tc}), with our definition of $\langle \rho_E\rangle$,  $\langle\mathcal{J}_z\rangle$ and  $\langle\rho_a\rangle$ (given above~(\ref{rhoEpf})-(\ref{rhoaf})) no dependence on the radial coordinate appears.
 
 
By using the fact that $\int_{-1}^1 d\xi \zeta(\xi) =1/2$ (which comes from the numerical calculation of the function $\zeta$ given in~\cite{Salvio:2025rma,dataset}) and setting rotation and chemical potentials to zero ($v=0$ and $\mathcal{M}^F_d=0$), one can explicitly check that for one Dirac fermion 
$$\langle \rho_E\rangle = 4\int \frac{d^3p}{(2\pi)^3} \omega f_F(\omega)$$ as it should be. 
 
\begin{figure}[t!!]
\begin{center}
  \includegraphics[scale=0.5]{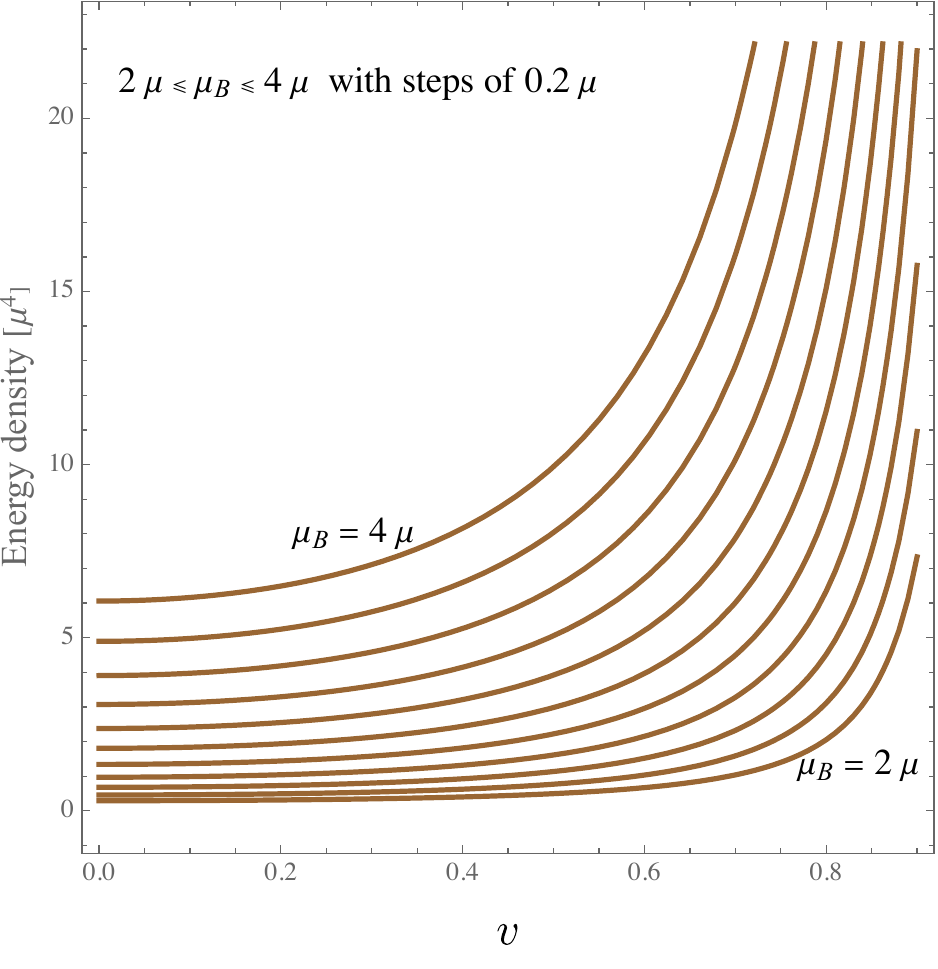}  \hspace{1cm} \includegraphics[scale=0.5]{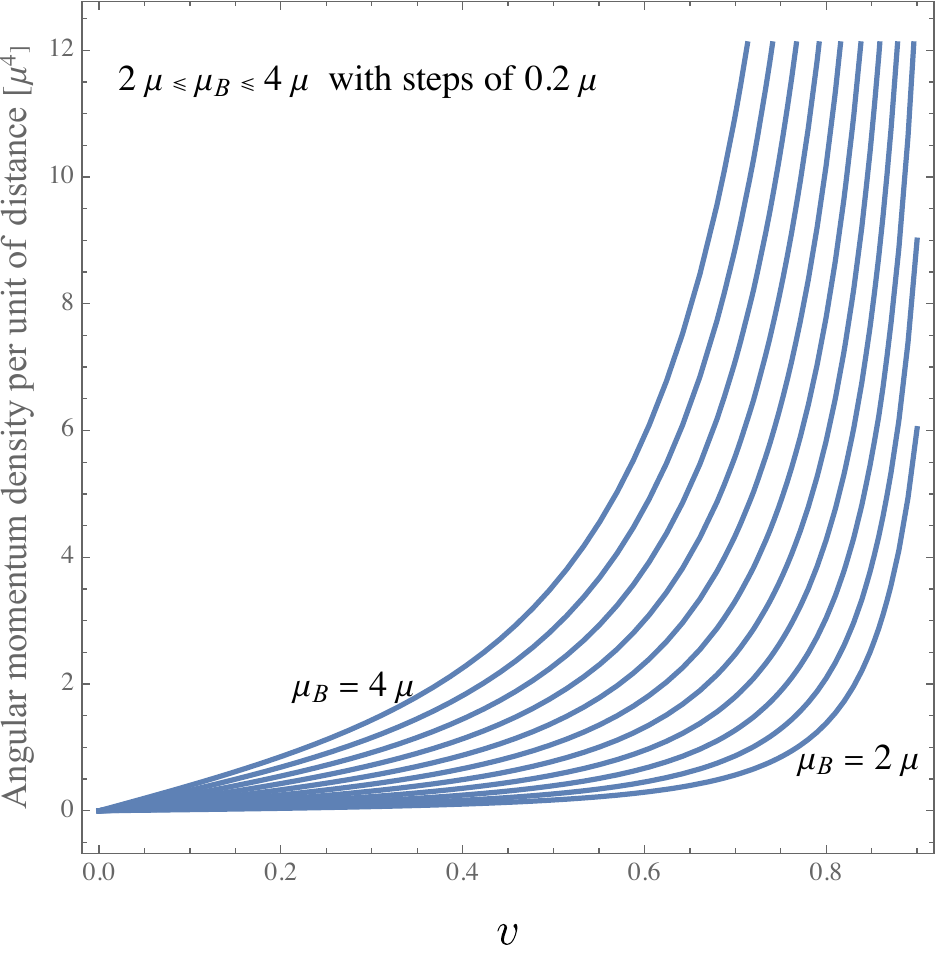} \\ 
 \includegraphics[scale=0.5]{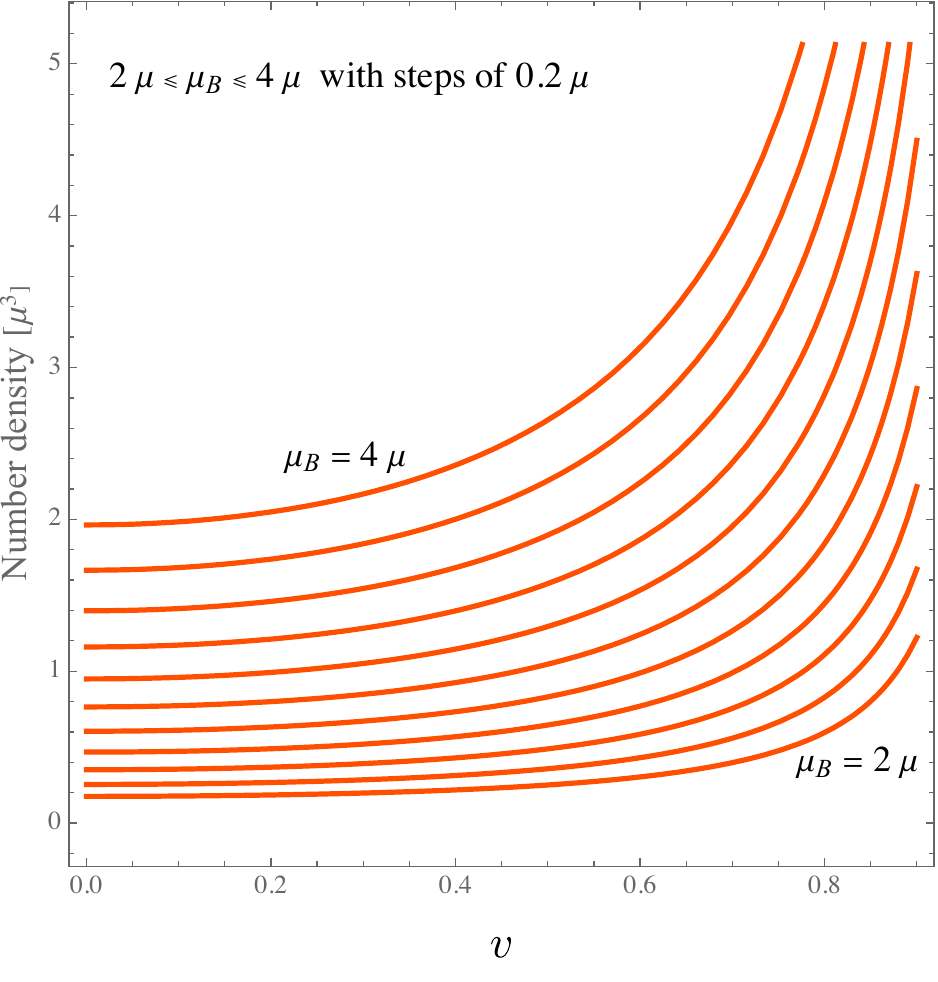}  \hspace{1cm} \includegraphics[scale=0.415]{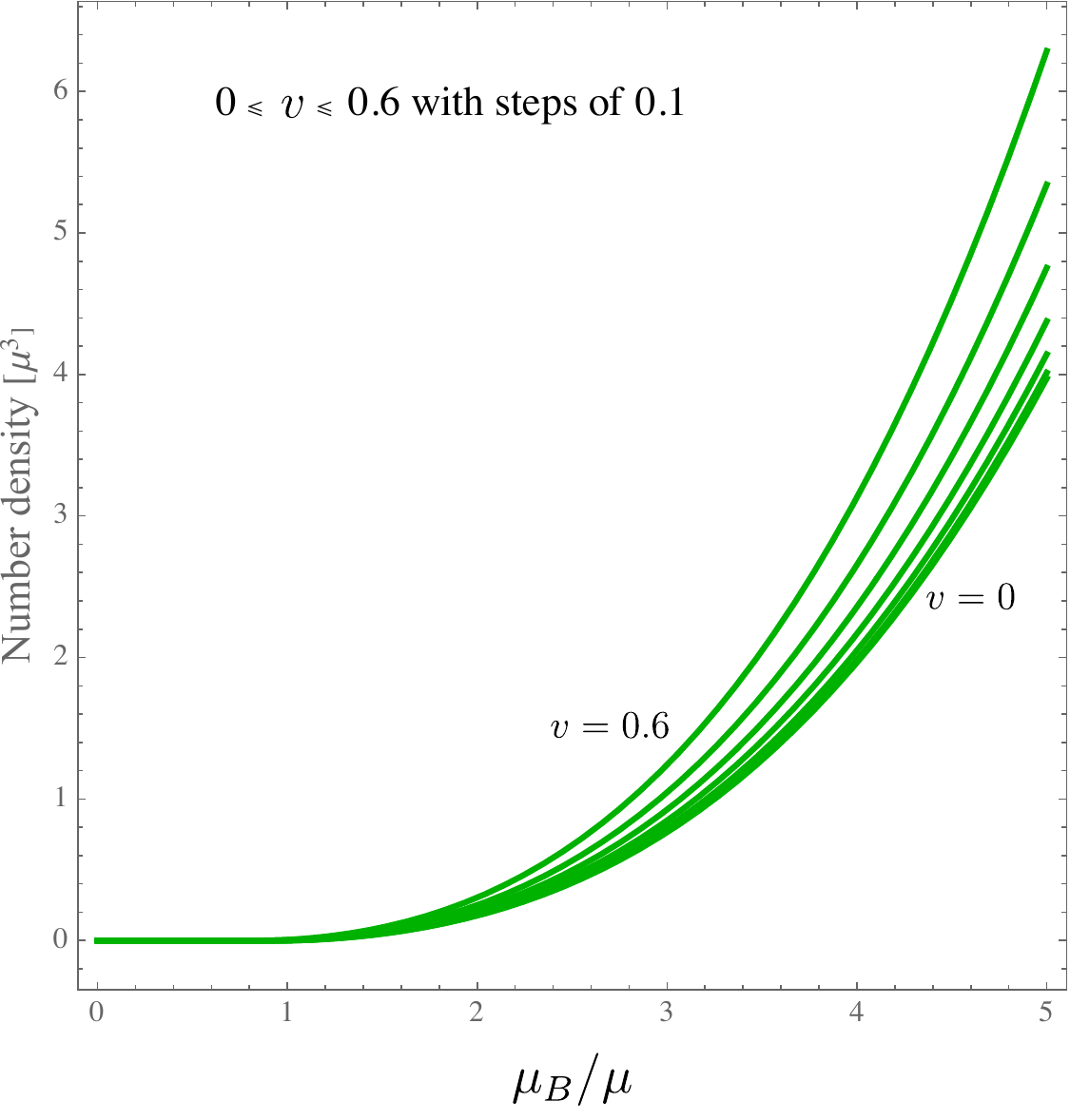}   \caption{\em Average energy density (upper left plot), average angular momentum density per unit of distance from the rotation axis (upper right plot) and average number density (lower left plot) as a function of the (rotational) velocity parameter $v$ in the case of a single Dirac fermion with mass $\mu$, a single chemical potential $\mu_B$ and $T\ll \mu$ (a relevant case for neutron stars). In the lower right plot it is shown how the average number density depends on $\mu_B$.}\label{rhoEJf}
  \end{center}
\end{figure}

Fig.~\ref{rhoEJf} shows $\langle \rho_E\rangle$, $\langle \mathcal{J}_z\rangle$ and $\langle\rho_a\rangle$ in the case of a single Dirac fermion with mass $\mu$ with a single chemical potential $\mu_B$ and $T\ll \mu$. This case is relevant for neutron stars\footnote{This may be true even  in merging- or proto-neutron stars~\cite{Naydenov:2025ydd}.}, identifying  $\mu_B$ with the effective baryon chemical potential and $\mu$ with the effective nucleon mass (see Sec.~\ref{Direct URCA processes in rotating neutron stars}). The lower right plot in Fig.~\ref{rhoEJf} tells us how to convert the chemical potential to the more physically transparent number density. Fig.~\ref{rhoEJf} clearly confirms our analytical proof that $\langle \rho_E\rangle$, $\langle \mathcal{J}_z\rangle$ and $\langle\rho_a\rangle$ become arbitrarily large as $v\to 1$.  Tables containing the numerical determination of the quantities plotted in Fig.~\ref{rhoEJf} can be found at~\cite{datasetf}.

%
%

A star of mass $M_s$ and radius $R_s$ that is held together only by gravitation can rotate up to a value of $\Omega$ of about $\Omega_{\rm max}\simeq \sqrt{G_N M_s/R_s^3}$, with $G_N$ being Newton's constant; so a neutron star at the Oppenheimer-Volkoff limit in~\cite{Oppenheimer:1939ne,WeinbergAstro}, $M_s\simeq 0.7M_{\rm sun}$, with
$R_s \simeq 10$\,km can reach a linear velocity at its surface of about $v_{\rm max}\simeq 0.32$~\cite{WeinbergAstro}. Using the up-to-date upper limit on the mass of rotating neutron stars  $M_s\simeq 2.6M_{\rm sun}$ obtained through gravitational-wave observations~\cite{Rezzolla:2017aly}, and $R_s \simeq 10$\,km, gives instead $v_{\rm max}\simeq 0.62$.
On the other hand, particles in the coronas of black holes can even orbit at velocities approaching the speed of light, $v_{\rm max}\simeq 1$.

Let us clarify now what changes in the case of Majorana fermions, Eq.~(\ref{FreeWexp}). Since in that case one has one type of annihilation operators, $a_{qs}$, rather than two, $c_{qs}$ and $d_{qs}$, only the contribution due to fermions  in~(\ref{rhofref}) is present, while that of antifermions  is absent. This is due to the fact that Majorana particles  coincide with their antiparticles. Correspondingly, $\langle H\rangle$, $\langle J_z\rangle$ and $\langle Q^a\rangle$ (and thus $\langle \rho_E\rangle$, $\langle \mathcal{J}_z\rangle$ and $\langle\rho_a\rangle$ too) will only have the fermion contribution, not that of antifermions.

\subsection{Thermal propagator}\label{Thermal propagator f}

The formul\ae~derived so far are useful, among many other things, to compute the thermal propagator, which plays a crucial
role in perturbation theory (some examples in the presence of $\Omega$ and $\mu_a$ will be studied in Secs.~\ref{Weakly-coupled fermion production} and~\ref{Direct URCA processes in rotating neutron stars}). 

For Dirac fermions this function is defined by
\be  \langle {\cal T} \Psi_s(x_1)\bar \Psi_{s'}(x_2)\rangle = \theta(t_1-t_2)\langle \Psi_s(x_1)\bar \Psi_{s'}(x_2)\rangle  -\theta(t_2-t_1)  \langle \bar \Psi_{s'}(x_2)\Psi_s(x_1)\rangle, \label{Tpropf}\ee
where the spinor indices are understood.

Using now the fermion field expansion in~(\ref{FreeFexp}) and the average of pairs of annihilation and creation operators in~(\ref{dgammad})-(\ref{deltadd}) one obtains for the non-time-ordered 2-point functions:
\bea &&S_{ss'}^>(x_1,x_2)  \equiv \langle \Psi_s(x_1)\bar \Psi_{s'}(x_2)\rangle \nonumber \\
&& \hspace{-1cm}=\SumInt_q \left\{\mathcal{U}_q(x_1)\mathcal{\bar U}_q(x_2)(1-f_F(\omega - (m+1/2)\Omega - \mu_a \bar {\cal R}^a))_{s's}+\mathcal{V}_q(x_1)\mathcal{\bar V}_q(x_2)f_F(\omega - (m+1/2)\Omega - \mu_a  {\cal R}^a)_{ss'}\right\} \nonumber\\ 
&&S_{ss'}^<(x_1,x_2) \equiv -\langle \bar \Psi_{s'}(x_2)\Psi_s(x_1)\rangle\nonumber \\ 
&&\hspace{-1cm}=-\SumInt_q \left\{\mathcal{U}_q(x_1)\mathcal{\bar U}_q(x_2) f_F(\omega - (m+1/2)\Omega - \mu_a \bar {\cal R}^a)_{s's}+\mathcal{V}_q(x_1)\mathcal{\bar V}_q(x_2)(1-f_F(\omega - (m+1/2)\Omega - \mu_a  {\cal R}^a))_{ss'}\right\} \nonumber \eea
that gives us the thermal propagator through Eq.~(\ref{Tpropf}). However, in computing particle decays or production, like in the examples of Secs.~\ref{Weakly-coupled fermion production} and~\ref{Direct URCA processes in rotating neutron stars}, it is sometimes easier to work with the non-time-ordered 2-point functions. Also, note that in the case of massless fermions that have just one helicity state, only one term in the sum over helicities in the expressions above should be selected.

Let us now consider the case of Majorana fermions, which are described here by Weyl spinors,  Eq.~(\ref{FreeWexp}). Then we can consider the following types of thermal propagators
 \bea  \langle {\cal T} \Psi_s(x_1)\bar \Psi_{s'}(x_2)\rangle &=& \theta(t_1-t_2)\langle \Psi_s(x_1)\bar \Psi_{s'}(x_2)\rangle  -\theta(t_2-t_1)  \langle \bar \Psi_{s'}(x_2)\Psi_s(x_1)\rangle,  \label{TpropfW1} \\ 
 \langle {\cal T} \Psi_s(x_1)\Psi_{s'}(x_2)\rangle &=& \theta(t_1-t_2)\langle \Psi_s(x_1)  \Psi_{s'}(x_2)\rangle  -\theta(t_2-t_1)  \langle \Psi_{s'}(x_2)\Psi_s(x_1)\rangle, \label{TpropfW2} \eea
where the spinor indices are understood. 
The field expansion in~(\ref{FreeWexp}) then leads to
\bea &&S_{ss'}^>(x_1,x_2)  \equiv \langle \Psi_s(x_1)\bar \Psi_{s'}(x_2)\rangle \nonumber \\
&&\hspace{-1cm}= \SumInt_q \left\{X_q(x_1)\bar X_q(x_2)(1-f_F(\omega - (m+1/2)\Omega - \mu_a  {\cal R}^a))_{s's}+Y_q(x_1)\bar Y_q(x_2)f_F(\omega - (m+1/2)\Omega - \mu_a  {\cal R}^a)_{ss'}\right\} \nonumber\\ 
&&S_{ss'}^<(x_1,x_2) \equiv -\langle \bar \Psi_{s'}(x_2)\Psi_s(x_1)\rangle\nonumber \\ 
&&\hspace{-1cm}=-\SumInt_q \left\{X_q(x_1)\bar X_q(x_2) f_F(\omega - (m+1/2)\Omega - \mu_a  {\cal R}^a)_{s's}+Y_q(x_1)\bar Y_q(x_2)(1-f_F(\omega - (m+1/2)\Omega - \mu_a  {\cal R}^a))_{ss'}\right\} \nonumber \\  
&&\tilde S_{ss'}^>(x_1,x_2)  \equiv \langle \Psi_s(x_1)  \Psi_{s'}(x_2)\rangle \nonumber \\
&&\hspace{-1cm}= \SumInt_q \left\{X_q(x_1)Y_q(x_2)(1-f_F(\omega - (m+1/2)\Omega - \mu_a  {\cal R}^a))_{s's}+Y_q(x_1)X_q(x_2)f_F(\omega - (m+1/2)\Omega - \mu_a  {\cal R}^a)_{ss'}\right\},\nonumber\eea
while $\tilde S_{ss'}^<(x_1,x_2)  \equiv -\langle \Psi_{s'}(x_2)  \Psi_{s}(x_1)\rangle$ 
can easily be obtained from $\tilde S_{ss'}^>(x_1,x_2)$ by exchanging $s\leftrightarrow s'$ and $x_1\leftrightarrow x_2$. 

If it turns out to be convenient, one can easily combine the non-time-ordered 2-point functions to obtain the thermal propagators in~(\ref{Tpropf}),~(\ref{TpropfW1}) and~(\ref{TpropfW2}) expressed in terms of an integro-sum over four momentum variables  $\{k_0, \alpha, p, m\}$. This can be done by  integrating over $\alpha$ and $p$  instead of $\omega= \sqrt{\mu^2+\alpha^2 +p^2}$ and $p$ (using the Jacobian formula $d\omega dp=\alpha d\alpha dp/\omega$) in the domain $\alpha\in[0,\infty)$ and $p\in(-\infty, +\infty)$ and  the following integral representation of the Heaviside step function
\be \theta(t) = \lim_{\varepsilon\to 0^+}\frac1{2\pi i} \int_{-\infty}^{+\infty} \frac{dk_0}{k_0 -i\varepsilon} e^{ik_0 t}.  \label{4Mom}\ee

To the best of our knowledge an explicit closed-form expression for the thermal fermion propagator (defined as in this work) was never obtained before in the presence of $\Omega$ and at finite temperature. Note that here not only $\Omega$, but also an arbitrary number of chemical potentials is included.  Let us remark,  however, that in~\cite{Vilenkin:1980zv} a different 2-point function was discussed for Dirac fermions: the thermal average of the time-ordered product  of the ``Matsubara field operator". This operator does not coincide with $\Psi$ but with the field $\tilde \Psi$, which will be defined in the general Eq.~(\ref{tildeOf}), after the substitution  $it\to \tau$, where $\tau$ is the imaginary time. In the absence of the $\mu_a$, Ref.~\cite{Vilenkin:1980zv} related this 2-point function to the known expression for $\Omega=0$ through the action of a differential operator, both in coordinate and momentum space. Another  connected work is Ref.~\cite{Ayala:2021osy} where, in some approximation, the fermion propagator for  $T=0$ and $\mu_a=0$ was computed in a rotating environment in momentum space.

\section{Fermion path integral}\label{Fermion path integral}
So far free fermions or quasi-free fermions (where in-medium effects are included by effective masses  and effective chemical potentials) have been discussed. Now, in order to investigate general fermion interacting theories, with arbitrary values of $\Omega$, as well as $T$  and the $\mu_a$, path integral methods are used.
Here it is obtained the path-integral formula for the thermal Green's functions
   \be  \langle \mathcal{T} \mathcal{O}_1(x_1) ...  \mathcal{O}_n(x_n) \rangle = \Tr (\rho \mathcal{T} \mathcal{O}_1(x_1) ...  \mathcal{O}_n(x_n)), \label{GreenO} \ee
   where the $\mathcal{O}_i$ are 
   operators involving the fermion fields $\Psi_s$. Such a formula, in the presence of $\Omega$ and $\mu_a$ can be derived by combining the corresponding discussion in the absence of $\Omega$ and $\mu_a$ of Ref.~\cite{Salvio:2024upo} with the derivation of the path-integral formula of thermal Green's functions for purely scalar theories of~\cite{Salvio:2025rma}, as it is now illustrated.
   
   First, one groups together all fermion annihilation operators ($c_{qs}$ and $d_{qs}$ for Dirac fermions and $a_{qs}$ for Majorana fermions  in~(\ref{FreeWexp})) in a single $a_i$. Here $i$ is a collective index running over both particles and antiparticles (if they are distinct from the particles) as well as all values of $q$ and $s$. One can then introduce right, $|\eta\rangle$, and left, $\langle\eta|$, ``eigenstates" of $a_i$ and $a_i^\dagger$, respectively: 
   \be a_i |\eta\rangle = \eta_i |\eta\rangle, \qquad \langle\eta| a_i^\dagger =  \langle\eta|\eta_i^*, \ee
   where $\eta_i$ and $\eta_i^*$ are Grassmann variables satisfying
   \be \{\eta_i,\eta_j\} =\{\eta_i^*,\eta_j^*\} =\{\eta_i,\eta_j^*\}  = 0, \quad \{\eta_i,a_j\}= \{\eta_i,a^\dagger_j\}=  \{\eta^*_i,a_j\} = \{\eta^*_i,a^\dagger_j\}= 0.  \ee

One can then express the trace of an observable (or more generally an operator commuting with the $\eta_i$ and the $\eta_i^*$) $A$ through the well-known formula (see~\cite{Salvio:2024upo} for all details)
 \be  \Tr (A)=\int d\eta^*d\eta \, e^{-\eta^*\eta} \langle -\eta|A|\eta\rangle, \label{TrFer} \ee
where $d\eta^* d\eta \equiv \prod_i d\eta_i^*d\eta_i$ and, for two generic values $\eta_i$ and $\eta_i'$ of the Grassmann variables, $\eta^*\eta'\equiv \sum_i \eta^*_i\eta'_i$. Instead of computing the trace through the states $|\eta\rangle$ one can equivalently use 
  \be |\tilde\eta,t\rangle \equiv \exp(i(H-\vec\Omega\cdot \vec J\,)t) |\eta\rangle. \label{tildeeta}\ee
  The tilde on $\eta$ distinguishes these states from $|\eta,t\rangle \equiv \exp(iHt) |\eta\rangle$, which are used in the standard derivation of the fermionic path-integral formula for $\Omega=0$ and $\mu_a=0$. Note that the states $|\tilde\eta,t\rangle$ and $\langle\tilde\eta,t|$ are right and left ``eigenstates" of 
  \bea \tilde a_i(t) &\equiv& \exp(i(H-\vec\Omega\cdot \vec J\,)t) a_i \exp(-i(H-\vec\Omega\cdot \vec J\,)t), \\   \tilde a_i^\dagger(t)&\equiv& \exp(i(H-\vec\Omega\cdot \vec J\,)t) a_i^\dagger\exp(-i(H-\vec\Omega\cdot \vec J\,)t), \eea 
respectively,  in the sense that 
   \be \tilde a_i(t) |\tilde\eta,t\rangle = \eta_i |\tilde\eta,t\rangle, \qquad \langle\tilde\eta,t|  \tilde a_i^\dagger(t) =  \langle\tilde\eta,t|\eta_i^*. \label{EigEq} \ee
   Correspondingly, regarding the operators $\mathcal{O}_i$, we define
  \be \mathcal{\tilde O}_i(t,\vec x) \equiv  \exp(i(H-\vec\Omega\cdot \vec J\,)t)\,\mathcal{O}_i(0,\vec x)\,\exp(-i(H-\vec\Omega\cdot \vec J\,)t) =  \mathcal{D}_i(t\vec\Omega)\mathcal{O}_i(t, R^{-1}(t\vec\Omega)\vec x), \label{tildeOf}\ee
 where $\mathcal{D}_i(t\vec\Omega)$  implements the rotation of angle $t\Omega$ around $\vec\Omega$ in the Lorentz representation of $\mathcal{O}_i$ and $R(t\vec\Omega)$ is the $3\times3$ orthogonal matrix with unit determinant representing the rotation associated with $t\vec\Omega$.
 For the trace one finds (starting from~(\ref{TrFer}) and using the ciclicity of the trace)
  \be  \Tr (A)=\int d\eta^*(t_0)d\eta(t_0) \, e^{-\eta^*(t_0)\eta(t_0)} \langle -\tilde\eta,t_0|A|\tilde\eta,t_0\rangle, \label{TrFert} \ee
  where $t_0$ is an arbitrary reference time. We have also labeled  with $(t_0)$ the integration variables  for subsequent notational convenience. 
  
 Applying~(\ref{TrFert}) to~(\ref{GreenO}) one obtains
  \bea \langle {\cal T}\mathcal{O}_1(x_1) ...  \mathcal{O}_n(x_n) \rangle= \int d\eta^*(t_0)d\eta(t_0) \, e^{-\eta^*(t_0)\eta(t_0)} \langle-\tilde\eta,t_0|\rho\,  {\cal T}\mathcal{O}_1(x_1) ...  \mathcal{O}_n(x_n)|\tilde\eta,t_0\rangle \nonumber \\
=\frac1{Z}\int d\eta^*(t_0)d\eta(t_0) \, e^{-\eta^*(t_0)\eta(t_0)} \langle-\tilde\eta,t_0|e^{-\beta (H-\vec\Omega  \cdot \vec J - \mu_a Q^a)}  {\cal T}\mathcal{O}_1(x_1) ...  \mathcal{O}_n(x_n)|\tilde\eta,t_0\rangle. \label{FermionTGF}\eea 
From the definition of the $|\eta,t\rangle$ one can write
\be \langle-\tilde\eta,t_0|e^{-\beta (H-\vec\Omega  \cdot \vec J - \mu_a Q^a)} = \langle-\exp(\beta\mu_aT^a)\tilde\eta,t_0-i\beta| \label{rhoOnS}\ee
as, by definition, the matrices $T^a$ act on the species index of the $a_i$ (and thus of the $\eta_i$) as follows:
\be \exp(i\alpha_aQ^a)a\exp(-i\alpha_aQ^a) = \exp(i\alpha_a T^a)a.  \ee 
 In other words, the $T^a$ are the generators of $\mathcal{G}$ in the representation of the $a_i$ (and thus of the $\eta_i$). In the case of Majorana fermions, for which $t^a=\bar t^a\equiv-(t^a)^*$, one takes $T^a=t^a$, while for Dirac fermions 
one can write the $T^a$ in a box-diagonal form, $T^a = $ diag$(t^a,\bar t^a)$, where the first block corresponds to the particles and the second one to the antiparticles. Using~(\ref{rhoOnS}) in~(\ref{FermionTGF}) leads to
\hspace{-1cm}\bea  \langle {\cal T}\mathcal{O}_1(x_1) ...  \mathcal{O}_n(x_n) \rangle=\frac1{Z}\int d\eta^*(t_0)d\eta(t_0) \, e^{-\eta^*(t_0)\eta(t_0)} \langle-\exp(\beta\mu_aT^a)\tilde\eta,t_0-i\beta| {\cal T}\mathcal{O}_1(x_1) ...  \mathcal{O}_n(x_n)|\tilde\eta,t_0\rangle \nonumber \\ 
=\frac1{Z}\int d\eta^*(t_0)d\eta(t_0) \, e^{-\eta^*(t_0)\eta(t_0)} \langle-\exp(\beta\mu_aT^a)\tilde\eta,t_0-i\beta| {\cal T}\mathcal{\tilde O}^\omega_1(x_1^\omega) ...  \mathcal{\tilde O^\omega}_n(x_n^\omega)|\tilde\eta,t_0\rangle, \nonumber \eea 
where $x^\omega\equiv \{t, R(t\vec\Omega)\vec x\}$ and the operators $\mathcal{\tilde O}^\omega_i(x_i^\omega)$ are defined by   $\mathcal{\tilde O}^\omega_i(x_i^\omega)\equiv \mathcal{D}^{-1}_i(t_i\vec\Omega)\mathcal{\tilde O}_i(x_i^\omega)$. Inverting~(\ref{tildeOf}) one actually finds $\mathcal{\tilde O}^\omega_i(x_i^\omega) = \mathcal{O}_i(x_i)$~\cite{Salvio:2026ewl}. The reason why we rewrite the $\mathcal{O}_i(x_i)$ in this way is to have all operators expressed in terms of the $\tilde a_i(t)$ and  $\tilde a^\dagger_i(t)$ rather than the $a_i$ and the $a_i^\dagger$, which allows us to use the ``eigenvalue" equations in~(\ref{EigEq}).

Then the rest of the derivation 
is the same as the well-known one in the absence of $\Omega$ and  the $\mu_a$ (see e.g.~Ref.~\cite{Salvio:2024upo} for details) except that here one has $H-\vec\Omega\cdot \vec J$ in place of $H$ and one uses the states $|\tilde\eta,t\rangle$ instead of $|\eta,t\rangle$ and the expression $\mathcal{\tilde O}^\omega_i(x_i^\omega)$ rather than $\mathcal{O}_i(x_i)$. So one finds 
\be
\hspace{-0.6cm} 
\boxed{\hspace{-0.1cm}\langle \mathcal{T}\mathcal{O}_1(x_1) ... \mathcal{O}_n(x_n)\rangle 
\hspace{-0.1cm}= \hspace{-0.1cm}\frac1{Z} \int \delta\eta^*\delta\eta \exp\left(\int_C dt\left(-\eta^*(t)\dot \eta(t)-iH^\omega_c(\eta^*(t),\eta(t)\right) )\right)O_1^{\omega}(x_1^\omega) ... O^{\omega}_n(x_n^\omega),} \hspace{-0.4cm} \label{PIGFF} 
\ee
where 
\be H^\omega_c(\eta^*,\eta') \equiv \frac{\langle\eta|H^\omega(a^\dagger,a) |\eta' \rangle}{\langle\eta|\eta' \rangle}, \quad  \mbox{with} \quad H^\omega \equiv H-\vec\Omega\cdot \vec J \ee
and the $O_i(x_i)$ are the $c$-number fields obtained by substituting (in the field operators $\mathcal{O}_i(x_i)$) the annihilation and creation operators with  $\eta$ and $\eta^*$, respectively, after putting all annihilation operators on the right of all creation operators. The definition $O_i^{\omega}(x_i^\omega) \equiv \mathcal{D}^{-1}_i(t_i\vec\Omega)O_i(x_i^\omega)$ has been used too.
  The measure in~(\ref{PIGFF}) is 
\be\delta\eta^*\delta\eta \equiv \prod_t d\eta^*(t) d\eta(t). \ee 
Also,  in the path integral, while the space integral has  no restriction,
 the integral over $t$ is performed on a contour $C$ in the complex $t$ plane that connects the arbitrary time  $t_0$ and $t_0-i\beta$ and contains the time components $x_1^0, ... , x_n^0$ of   $x_1, ... , x_n$. The arbitrariness of $t_0$ allows us to adopt the real- or the imaginary-time formalism by choosing $C$ appropriately.
The partition function $Z$ in~(\ref{PIGFF}) is just the numerator in~(\ref{PIGFF}) for $O_1(x_1) ... O_n(x_n)\to 1$:
\be
\hspace{-0.9cm} 
\boxed{Z
=  \int \delta\eta^*\delta\eta \exp\left(\int_C dt\left(-\eta^*(t)\dot \eta(t)-iH^\omega_c(\eta^*(t),\eta(t))\right) \right).} \hspace{-0.5cm} \label{PIZF} 
\ee
This expression is useful as $Z$ can be used to compute the averages of observables, as explained in general terms in~\cite{Salvio:2025rma}.
The integration in~(\ref{PIGFF}) and~(\ref{PIZF}) is subject to the twisted antiperiodic boundary conditions, 
\be \eta(t_0-i\beta)=-\exp(\beta\mu_aT^a)\eta(t_0), \quad  \eta^*(t_0-i\beta)=-\exp(\beta\mu_a\bar T^a)\eta^*(t_0), \label{ABC}\ee
where $\bar T^a \equiv -(T^a)^*$. These conditions are direct consequences of~(\ref{rhoOnS}). The formula in~(\ref{PIGFF}) represents the path-integral formula for general operators $\mathcal{O}_1, ... ,\mathcal{O}_n$ involving fermion fields, whose dynamics is dictated by an arbitrary Hamiltonian.

 Like in the purely scalar theories, to account for a non-vanishing average angular momentum one has to substitute  
  $H_c$ with $H_c -\vec \Omega\cdot \vec J_c$ in the path integral, which is nothing but the transformation rule of the classical Hamiltonian from an inertial frame to a frame rotating with angular-velocity vector $\vec \Omega$.      One can, therefore, identify $\vec \Omega$ with the angular-velocity vector of a rigidly-rotating plasma.

Adapting again the discussion of Ref.~\cite{Salvio:2024upo} (done with $\Omega=0$ and $\mu_a=0$) to the presence of general $\Omega$ and $\mu_a$, the path integral in~(\ref{PIGFF}) can then be written in terms of $c$-number Grassmann fermion fields  $\psi$ and $\bar \psi$ as follows\footnote{For Dirac fermions one can easily show~(\ref{PIGFFFo}) through the orthogonality conditions in~(\ref{OrtoF}). For the Majorana fermions in~(\ref{FreeWexp}) the proof is complicated by the fact that in general the orthogonality conditions in~(\ref{OrtoFX}) and~(\ref{OrtoFY}) do not include the vanishing of $\int d^3 x \, \bar Y_{q'}(x)X_{q}(x)$. Indeed, this integral is non zero for some $q$ and $q'$  because the $X_q$ and $Y_{q}$ can be viewed as two different basis in the same linear space of functions. In particular it can be non zero for $\sigma'=\sigma$, $p'=-p$, $m'+1/2=-m-1/2$ and $|\vec p'|=|\vec p|$. However, the non-vanishing of that integral produce the following terms in $\int_C d^4 x \,   \psi^\dagger(x)\dot\psi(x)$:
\be \SumInt_{q'}\SumInt_q\int_C d^4x\left[\bar X_{q'}(0,\vec x) Y_q(0,\vec x) \alpha^*_{q'}\dot\alpha^*_q+\bar Y_{q'}(0,\vec x) X_q(0,\vec x) \alpha_{q'}\dot\alpha_q\right], \label{ExtraPI}\ee 
where the $\alpha_q$ and $\alpha_q^*$ are the Grassmann variables corresponding to $a_{qs}$ and $a^\dagger_{qs}$, respectively, with species index $s$ understood. Now,~(\ref{YfromX}) (together with the conditions $\sigma'=\sigma$ and $|\vec p'|=|\vec p|$) implies that $\bar X_{q'}(0,\vec x) Y_q(0,\vec x)$ and $\bar Y_{q'}(0,\vec x) X_q(0,\vec x)$ are antisymmetric in $q\leftrightarrow q'$, while an integration by parts over time shows that $\int_Cdt\alpha^*_{q'}\dot\alpha^*_q$ and $\int_Cdt\alpha_{q'}\dot\alpha_q$ are symmetric because $\alpha_q$ and $\alpha_q^*$ are  Grassmann variables. So~(\ref{ExtraPI}) vanishes.}
\bea  &&\hspace{-1cm} \langle \mathcal{T}\mathcal{O}_1(x_1) ... \mathcal{O}_n(x_n)\rangle \nonumber \\
&&\hspace{-1cm}= \frac1{``O_i\to 1"} \int \delta\bar\psi\delta\psi\exp\left(\int_C d^4x\left[-\psi^\dagger(x)\dot \psi(x)-i\mathcal{H}^\omega_c(\bar\psi(x),\psi(x))\right] \right)O_1^\omega(x_1^\omega) ... O_n^\omega(x_n^\omega), \label{PIGFFFo} \eea
subject to the antiperiodic boundary conditions,
\be \psi(t_0-i\beta,\vec x)=-\exp(\beta\mu_at^a)\psi(t_0,\vec x), \quad \bar \psi(t_0-i\beta,\vec x)=-\bar\psi(t_0,\vec x)\exp(-\beta\mu_at^a), \label{AntiPpsi}\ee
where $\bar \psi\equiv \psi^\dagger \gamma^0$ for Dirac spinors and $\bar \psi\equiv \psi^\dagger$ for Weyl spinors.
Here, in~(\ref{PIGFFFo}),  $\mathcal{H}^\omega_c$ is the full classical Hamiltonian density, including the effect of rotation. The defining property of this quantity is
  \be \int d^3x \, \mathcal{H}^\omega_c = H_c^\omega =-\vec\Omega\cdot\vec J_c+ \int d^3x \, \mathcal{H}_c, \label{callHc2}\ee
  where $\vec J_c$ is defined through 
  \be \vec J_c(\eta^*,\eta') \equiv \frac{\langle\eta| \vec J(a^\dagger,a) |\eta' \rangle}{\langle\eta|\eta' \rangle} \ee
  and 
  $\mathcal{H}_c$ is the corresponding classical Hamiltonian density in the absence of rotation. For Dirac spinors 
  \be \vec J_c = \int d^3x \psi^\dagger(x)\left[\vec x\times(-i\vec\nabla)+\frac{\vec\sigma_4}{2}\right]\psi(x),\label{DiracJc}\ee 
  where the three components of $\vec\sigma_4$ are the following $4\times 4$ matrices,
  \be \vec\sigma_4 = \left(\bac \sigma^{23} \\ \sigma^{31}\\ \sigma^{12} \ea \right), \quad \mbox{with} \quad \sigma^{\mu\nu} \equiv\frac{i}2 [\gamma^\mu,\gamma^\nu]. \ee 
  For Weyl spinors
   \be \vec J_c = \int d^3x \psi^\dagger(x)\left[\vec x\times(-i\vec\nabla)+\frac{\vec\sigma}{2}\right]\psi(x). \label{WeylJc}\ee
Then, one can write 
\bea  &&\hspace{-1cm} \langle \mathcal{T}\mathcal{O}_1(x_1) ... \mathcal{O}_n(x_n)\rangle \nonumber \\
&&\hspace{-1cm}= \frac1{``O_i\to 1"} \int \delta\bar\psi\delta\psi\exp\left(i\int_C d^4x\, \mathscr{L}_\omega(\bar\psi(x),\psi(x))\right)O_1^\omega(x_1^\omega) ... O_n^\omega(x_n^\omega), \label{PIGFFF} \eea
where the full Lagrangian density including the effect of rotation, $\mathscr{L}_\omega$, is 
\be \mathscr{L}_\omega = \mathscr{L} + \psi^\dagger(x)\,\,\vec\Omega\cdot\left[\vec x\times(-i\vec\nabla)+\frac{\vec\sigma_c}{2}\right]\psi(x), \ee 
with $\mathscr{L}$ being the full Lagrangian density in the absence of rotation and $\vec\sigma_c=\vec\sigma_{4}$ for Dirac spinors and $\vec\sigma_c=\vec\sigma$ for Weyl spinors.

Like for purely scalar theories, $\vec\Omega$ only appears in the  quadratic action, which implies that, in perturbation theory, only the propagators are modified by $\vec\Omega$, the vertices are unmodified. The vertices are also unmodified by the $\mu_a$. For the computation of the vertices one can then use well-known results from the literature (see e.g.~\cite{Bellac:2011kqa} and~\cite{Landsman:1986uw}); on the other hand, the propagators have been computed, including the effect of $\vec\Omega$ and  $\mu_a$,  in Sec.~\ref{Thermal propagator f}.

As usual these Green's functions can be obtained by taking functional derivatives of
\be \mathcal{Z}(\bar\kappa,\kappa) = \frac1{``\{\bar\kappa,\kappa\}\to0"} \int \delta\bar\psi\delta\psi
 \exp\left(i\int_C d^4x\, \mathscr{L}_\omega(\bar\psi(x),\psi(x))+i\int_Cd^4x(\bar\kappa(x)\psi^\omega(x^\omega)+\bar\psi^\omega(x^\omega)\kappa(x))\right) \nonumber\ee 
 with respect to the Grassmann sources $\kappa$ and $\bar\kappa$. 
 
 Now one can easily combine the results of~\cite{Salvio:2025rma} for scalars with those of the present section to obtain the generating functional $\mathcal{Z}$ for a general scalar-fermion theory with arbitrary values of temperature, chemical potentials and average angular momentum:
  \bea  \mathcal{Z} (j,\bar\kappa,\kappa) =\frac1{``\{j, \bar\kappa,\kappa\}\to0"} \int \delta\varphi \, \delta p_\varphi  \delta\bar\psi\delta\psi\, \exp\left(i \int_C d^4x \left(\dot\varphi(x)p_\varphi(x)+i\psi^\dagger(x)\dot \psi(x) \right.\right. \nonumber \\
  \left.\left. - \mathcal{H}^\omega_c(\varphi(x), p_\varphi(x),\bar\psi(x),\psi(x))+j(x)\varphi(x^\omega)+\bar\kappa(x)\psi^\omega(x^\omega)+\bar\psi^\omega(x^\omega)\kappa(x)\right)\right), \nonumber\label{GenFunSF}\eea     
where now  the  classical Hamiltonian density that includes the effect of rotation, $\mathcal{H}^\omega_c$, takes into account both the scalar and the fermion contributions. 

As already mentioned, to obtain the path integral with $\vec \Omega\neq0$ one can start from the path integral with $\vec \Omega=0$ and substitute there the classical Hamiltonian in the non-rotating frame with that in the rotating frame with angular-velocity vector $\vec \Omega$.   As a result, if scalar fields are also introduced, the effective scalar Euclidean action obtained by integrating out these fermions must have a real part that is bounded from below when it is so for $\vec\Omega=0$. 
This extends the proof given in~\cite{Salvio:2025rma} for scalars to any bosonic theories, including those obtained by integrating out fermion fields.

  \section{Some applications}\label{Some applications}
  
  In this section we provide some applications of the general results obtained previously, paying special attention to cases of relevance for neutron stars. 
  
  \subsection{Fermi momentum and Fermi surface}\label{Fermi momentum and Fermi surface}
  
In Sec.~\ref{Computing ensemble averages f} it was shown that it is always possible to consider a basis where all fermion species have well-defined masses and chemical potentials, even if the internal symmetries are not Abelian. So let us call now $M_i$ and $\mu_i$ the (effective) 
mass and (effective) chemical potential of the  $
 i$-th fermion species, respectively. Here for the sake of definiteness we assume $\mu_i\geq 0$; taking instead $\mu_i<0$ would essentially switch the role of fermions and antifermions. In this section we determine the Fermi momentum and the Fermi surface in the presence of rotation. These quantities, which are introduced for strongly degenerate fermions, are useful, for example, to describe the physics of neutron stars.
 
Let us define the Fermi momentum of the $
 i$-th species as $P_{Fi}\equiv 
 \sqrt{\alpha_i^2+p^2_i}$ where $\alpha_i$ and $p_i$ are the maximal values of $\alpha$ and $p$ such that 
 \be \sqrt{
M^2_i+\alpha^2+p^2}-v \alpha \xi\leq \mu_i, \label{CondFrot}\ee
 in the integration domain of~(\ref{rhoEpf})-(\ref{rhoaf}) and for given values of $M_i$, $\mu_i$ and $v$. The condition in~(\ref{CondFrot}) comes from taking the low-temperature limit in the Fermi-Dirac distributions in~(\ref{rhoEpf})-(\ref{rhoaf}). To determine the explicit expression of $P_{Fi}$ let us first determine the condition on $\mu_i$ and $M_i$ for a fixed $v$ such that there exists actually a Fermi momentum. The left-hand side of~(\ref{CondFrot}) is larger than or equal to its value at $p=0$ and $\xi=1$. Setting $p=0$ and $\xi=1$, one finds that~(\ref{CondFrot}) has solutions with respect to $\alpha$ when
 \be \mu_i \geq \sqrt{1-v^2} M_i, \label{muFCond}\ee
 which extends the well-known  $\mu_i \geq M_i$ (valid at $v=0$) to finite values of $v\in [0,1)$. Interestingly, rotation  ($v\neq0$) favors the existence of a Fermi momentum.
For $p=0$ and $\xi=1$, Condition~(\ref{CondFrot}) then corresponds to
 \be \frac{v\mu_i-\sqrt{\mu_i^2-(1-v^2)M_i^2}}{ 1-v^2} \leq \alpha \leq \frac{v\mu_i+\sqrt{\mu_i^2-(1-v^2)M_i^2}}{ 1-v^2}, \qquad (\alpha\geq 0). \label{RangeAlpha}\ee
 Now, noting that taking $\xi=1$ allows us to maximise $\alpha$ and $p$ compatibly with~(\ref{CondFrot}), the maximal value of $p^2$ for which~(\ref{CondFrot}) is satisfied, is
 \be p^2_i(\alpha) = \mu_i^2-M_i^2-(1-v^2)\alpha^2 + 2v\mu_i\alpha, \ee  
 which is not negative if and only if~(\ref{RangeAlpha}) is satisfied. The Fermi momentum $P_{Fi}$ is then given by $\sqrt{\alpha^2+p^2_i(\alpha)}$ for the maximal value of $\alpha$ compatible with~(\ref{RangeAlpha}), which gives $p_i(\alpha)=0$ and so
 \be P_{Fi} = \frac{v\mu_i+\sqrt{\mu_i^2-(1-v^2)M_i^2}}{ 1-v^2}. \label{pFiv}\ee 
This is the expression for the Fermi momentum for generic values of $v\in[0,1)$. As it should, for $v=0$~(\ref{pFiv}) reduces to the well-known expression $P_{Fi} = \sqrt{\mu_i^2 - M_i^2}$. One can observe that the Fermi momentum $P_{Fi}$ grows with $v$ and becomes arbitrary large as $v$ approaches the speed of light.
 
 Note now that setting $p=p_i(\alpha)$ and $\alpha$ to the upper bound in~(\ref{RangeAlpha}), which gives $p_i(\alpha)=0$, one saturates the bound in~(\ref{CondFrot}) for $\xi=1$. This is a particular point on the Fermi surface defined by
 \be  \sqrt{
M^2_i+\alpha^2+p^2}-v \alpha \xi= \mu_i, \label{Fsurface1}\ee
or equivalently
\be p^2 =  -M_i^2-\alpha^2+(\mu_i+v\alpha\xi)^2=\mu_i^2-M_i^2 - (1-v^2\xi^2)\alpha^2+2v\mu_i\xi \alpha\geq0.\label{Fsurface2}\ee
The Fermi surface is the set of values of $\alpha$, $p$ and $\xi$ such that Eq.~(\ref{Fsurface1}) is satisfied. Setting $v=0$ one obtains the well-known expression $\alpha^2+p^2+M_i^2=\mu_i^2$. 

It is important to note that $P_{Fi}$ is not the only value of $\sqrt{\alpha^2+p^2}$ on the Fermi surface for $v\neq0$. For example, setting $\xi=0$ in Eq.~(\ref{Fsurface1}) one finds that another value of $\sqrt{\alpha^2+p^2}$ on the Fermi surface is, for any $v$, the Fermi momentum for non-rotating plasmas, $\sqrt{\mu^2_i-M_i^2}$. Also note that, since setting $p=0$ and $\xi=1$ allows us to maximise $\alpha$ compatibly with~(\ref{CondFrot}), the upper bound in~(\ref{RangeAlpha}) is also the maximal value of $\alpha$ on the Fermi surface. On the other hand, the maximal value of $|p|$ on the Fermi surface is $\sqrt{\mu_i^2/(1-v^2)-M_i^2}$, which can be obtained by maximizing the function of $\xi$ and $\alpha$ in~(\ref{Fsurface2}), i.e.~by setting $\xi=1$ and $\alpha = \mu_i v/(1-v^2)$. Both the maximal value of $\alpha$ and the maximal value of $p$ on the Fermi surface grow with $v$ and become arbitrary large as $v$ approaches the speed of light.

  \subsection{Weakly-coupled fermion production}\label{Weakly-coupled fermion production}
  
  We now turn to the production of a weakly-coupled fermion (or antifermion) described by a fermion field $\Psi$. Let us start from the production of a Dirac fermion, such as an electron or a Dirac neutrino. In the interaction picture $\Psi$ can be decomposed using the creation and annihilation operators, like in~(\ref{FreeFexp}). We consider an interaction between $\Psi$ and the thermal bath of the form $\lambda \bar\Psi O + \lambda^* \bar O \Psi$, where  $O$ is a local fermion operator made of the fields in thermal equilibrium and $\lambda$ is a small coupling constant.
  
  At leading order in $\lambda$, the $S$-matrix element for the  production of a fermion with a given eigenvalue, $q$, of $H$, $P^z$, $J_z$ and $\vec J\cdot \vec P/|\vec p|$  is
\be S_{if}(q)  \simeq i\lambda \int d^4x \, \langle f, q|\bar\Psi(x) O(x) |i\rangle = i\lambda \sqrt{\Delta\omega\Delta p} \int d^4x \,\mathcal{\bar U}_{q}(x) \langle f|O(x)|i\rangle, \label{S}\ee
where $|i\rangle$ and $|f,q\rangle$ are the initial and final states. 
The states $|i\rangle$ 
are chosen to be eigenstates of $H-\vec\Omega  \cdot \vec J - \mu_a Q^a$ with eigenvalues $\mathcal{E}_i$, such that the production probability averaged over the initial state and summed over $f$ is (at leading order in $\lambda$) 
\be \boxed{\frac1{Z}\sum_{if} e^{-\beta\mathcal{E}_i} |S_{if}(q)|^2 
= - \Delta\omega\Delta p \int d^4x_1 d^4x_2 \, \mathcal{\bar U}_{q\alpha}(x_1)\mathcal{U}_{q\beta}(x_2) \Sigma_{\alpha\beta}^<(x_1,x_2),}\label{probF}\ee 
where
\be 
  \Sigma_{\alpha\beta}^<(x_1,x_2) \equiv - |\lambda|^2\langle \bar O_\beta(x_2)O_\alpha(x_1)\rangle\equiv - \frac{|\lambda|^2}{Z}\sum_{i} e^{-\beta\mathcal{E}_i}\langle i|
\bar O_\beta(x_2)O_\alpha(x_1)|i\rangle \label{SigmaBackward}\ee
  is a non-time-ordered 2-point function of the operator $O$.
  
  Analogously, the $S$-matrix element for the production of the antifermion is, at leading order in $\lambda$, 
  \be \bar S_{if}(q)  \simeq  i\lambda^* \sqrt{\Delta\omega\Delta p} \int d^4x \, \langle f|\bar O(x)|i\rangle \mathcal{V}_{q}(x) \label{Sbar}\ee
  and so
  \be \boxed{\frac1{Z}\sum_{if} e^{-\beta\mathcal{E}_i} |\bar S_{if}(q)|^2 
=\Delta\omega\Delta p \int d^4x_1 d^4x_2 \, \mathcal{\bar V}_{q\alpha}(x_1)\mathcal{V}_{q\beta}(x_2)\Sigma_{\alpha\beta}^>(x_1,x_2),}\label{probAF}\ee 
where 
\be 
  \Sigma_{\alpha\beta}^>(x_1,x_2) \equiv |\lambda|^2\langle O_\alpha(x_1)\bar O_\beta(x_2)\rangle\equiv \frac{|\lambda|^2}{Z}\sum_{i} e^{-\beta\mathcal{E}_i}\langle i|
O_\alpha(x_1)\bar O_\beta(x_2)|i\rangle \label{SigmaForward}\ee
is another non-time-ordered 2-point function of the operator $O$. 

Let us now consider the production of a massive Majorana fermion. An example could be a sterile neutrino in a type-I  see-saw model.  Using the Weyl-spinor formalism, this time  $\Psi$ can be decomposed using the creation and annihilation operators like in~(\ref{FreeWexp}). The Majorana fermion has an interaction with the thermal bath of the form $\lambda \Psi O + \lambda^* \bar O \bar\Psi$ where  again $O$ is a local fermion operator made of the fields in thermal equilibrium and $\lambda$ is a small coupling constant.  At leading order in $\lambda$, the $S$-matrix element for the  production of a Majorana fermion with a given eigenvalue, $q$, of $H$, $P^z$, $J_z$ and $\vec J\cdot \vec P/|\vec p|$  is
\bea S_{if}(q)  &\simeq& i \int d^4x \, \langle f, q|\lambda \Psi(x) O(x) +\lambda^* \bar\Psi(x) \bar O(x)|i\rangle \\&=& i  \sqrt{\Delta\omega\Delta p} \int d^4x \left(\lambda Y_{q}(x) \langle f|O(x)|i\rangle+\lambda^* \bar X_{q}(x) \langle f|\bar O(x)|i\rangle\right). \label{SMaj}\eea
Thus, the production probability averaged over the initial state and summed over $f$ can be written (at leading order in $\lambda$) 
\bea  \frac1{Z}\sum_{if} e^{-\beta\mathcal{E}_i} |S_{if}(q)|^2 
=  \Delta\omega\Delta p \int d^4x_1 d^4x_2 \left( Y_{q\alpha}(x_1)\bar Y_{q\beta}(x_2) \Sigma_{1\alpha\beta}^<(x_1,x_2) \right. \nonumber \\ \left.+\bar X_{q\alpha}(x_1)X_{q\beta}(x_2) \Sigma_{2\alpha\beta}^<(x_1,x_2)+ Y_{q\alpha}(x_1)X_{q\beta}(x_2) \Sigma_{3\alpha\beta}^<(x_1,x_2)\right),\label{probFmaj}\eea
where 
\bea \Sigma_{1\alpha\beta}^<(x_1,x_2) &\equiv& |\lambda|^2\langle \bar O_\beta(x_2) O_\alpha(x_1)\rangle\equiv \frac{|\lambda|^2}{Z}\sum_{i} e^{-\beta\mathcal{E}_i}\langle i|
\bar O_\beta(x_2) O_\alpha(x_1)|i\rangle,     \\  
\Sigma_{2\alpha\beta}^<(x_1,x_2) &\equiv& |\lambda|^2\langle  O_\beta(x_2) \bar O_\alpha(x_1)\rangle\equiv \frac{|\lambda|^2}{Z}\sum_{i} e^{-\beta\mathcal{E}_i}\langle i|
O_\beta(x_2) \bar O_\alpha(x_1)|i\rangle,   \\  
\Sigma_{3\alpha\beta}^<(x_1,x_2) &\equiv& 
2\lambda^2\mbox{Re}\langle  O_\beta(x_2)  O_\alpha(x_1)\rangle\equiv 2\frac{\lambda^2}{Z}\sum_{i} e^{-\beta\mathcal{E}_i}\mbox{Re}\langle i|
O_\beta(x_2)  O_\alpha(x_1)|i\rangle \label{sigma123}
\eea 

It is important to note that the results in (\ref{probF})-(\ref{sigma123}), although only at leading order in $\lambda$, are valid to
all orders (and even non-perturbatively) in the couplings of the thermalized sector other than $\lambda$. 

 If perturbation theory holds, these non-time-ordered 2-point functions can be computed with the Kobes-Semenoff rules~\cite{KS,KS2}. In their work Kobes and Semenoff assumed $\vec\Omega = 0$ and $\mu_a = 0$, but, as shown in Sec.~\ref{Fermion path integral} (and in~\cite{Salvio:2025rma} for scalars), only the propagators are modified by $\vec\Omega$ and $\mu_a$, the vertices are unmodified. The propagators have been computed, including the effect of $\vec\Omega$ and $\mu_a$,  in Sec.~\ref{Thermal propagator f} (and in~\cite{Salvio:2025rma} for scalars).

  \subsection{Direct URCA processes in rotating neutron stars}\label{Direct URCA processes in rotating neutron stars}
  
  The direct URCA (DU) processes
\be n \to p +l^{-}+\bar\nu_l, \qquad p+l^{-}\to n  +\nu_l, \label{DirectUrca}\ee
when possible, are the leading cooling processes of neutron stars, several orders of magnitude more efficient than other neutrino-production processes. 
In~(\ref{DirectUrca}) $l$ can be an electron or a muon and $\nu_l$ is the corresponding neutrino. Ref.~\cite{Lattimer:1991ib} showed in the absence of rotation that the DU processes are active in neutron stars if the proton fraction $n_p/(n_n+n_p)$, where $n_i$ is the number density of the $
 i$-th species, is above a certain threshold. It is known that the spin down of rotating neutron stars, such as pulsars, may create the right conditions where the DU processes become operative~\cite{Negreiros:2011ak}.

 As shown in~\cite{Lattimer:1991ib}, the reason why the DU processes are blocked when the proton fraction is too low is because, at least for $v=0$, in order for those processes to be active, the sum of the Fermi momenta of protons, $P_{Fp}$, and leptons, $P_{Fl}$, should not be lower than  that of neutrons, $P_{Fn}$ (we assume typical temperatures of neutron stars for which the neutrino linear momentum is negligible). But the Fermi momenta are linked to the number densities of the corresponding species, and this, together with charge neutrality, leads generically to a lower bound on the number density of protons. The Fermi momentum and Fermi surface for a rotating plasma have been introduced in Sec.~\ref{Fermi momentum and Fermi surface}. As shown there, the Fermi momentum, which is the maximal value of $\sqrt{\alpha^2+p^2}$ on the Fermi surface, is not the only value of $\sqrt{\alpha^2+p^2}$ on the Fermi surface for $v\neq0$. Also, the relation between the Fermi momentum and the number density can be affected by $v$. So the condition for the DU processes to be active can be different turning on $v$ and it is thus interesting to consider this process in rotating neutron stars.

  The effective Lagrangian density for the DU processes in~(\ref{DirectUrca}) is (see Ref.~\cite{Gorchtein:2023srs} for an introduction to the neutron beta decay in the SM)
  \be \mathscr{L}_{l\nu p n}  = -\sqrt{2} G_F V_{ud} \left[\bar l \gamma_\mu \nu_{lL}\right]\left[ \bar p (g_V - g_A \gamma_5) \gamma^\mu n\right] + {\rm h.c.}\ee 
  having neglected the small nucleon recoil. 
   Here,   $\nu_{lL} = (1-\gamma_5)\nu_l/2$ is the left-handed neutrino field, $G_F$ is the Fermi constant and $V_{ud}$ is the top-left element of the Cabibbo-Kobayashi-Maskawa (CKM) matrix. 
   The couplings $g_V$ and $g_A$ are the  vector and axial couplings respectively.

\begin{figure}[t]
\begin{center}
\vspace{-3cm}
  \includegraphics[scale=0.52]{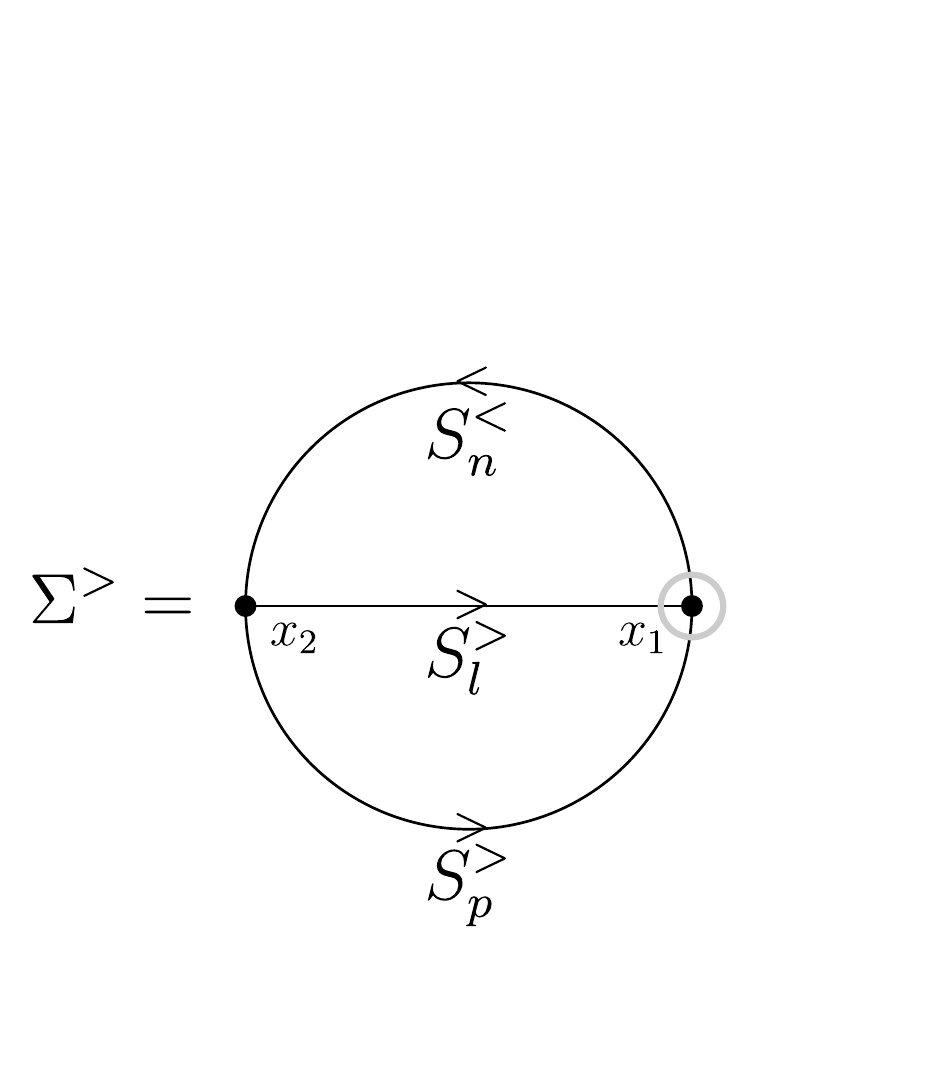}\quad 
  \includegraphics[scale=0.52]{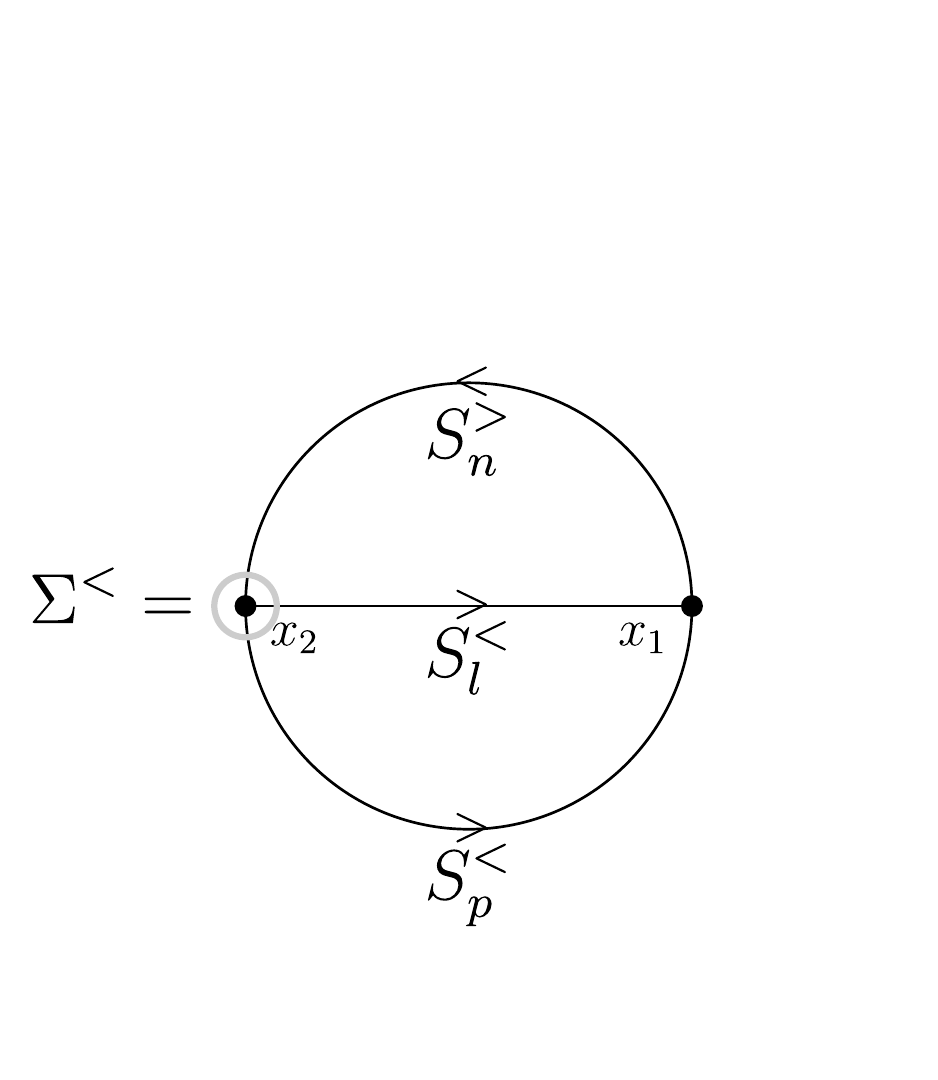}
      \caption{\em Diagrams representing the  non-time-ordered 2-point functions, Eq.~(\ref{SigmaForward}) on the left and Eq.~(\ref{SigmaBackward}) on the right, which are relevant for (anti)neutrino production via the DU processes in~(\ref{DirectUrca}) in terms of the  non-time-ordered 2-point functions of $n$, $p$ and $l$.  The Kobes-Semenoff circling notation~\cite{KS,KS2} is used.}\label{diagramF}
  \end{center}
\end{figure}

  Thus, in this case the operators $O$ and $\bar O$ of Sec.~\ref{Weakly-coupled fermion production} for neutrino and antineutrino production can be written as follows:
  \be O =   \gamma_\mu l_L  \left[ \bar n  (g_V - g_A \gamma_5) \gamma^\mu p\right],  \qquad \bar O = \left[ \bar p (g_V - g_A \gamma_5) \gamma^\mu n\right]\overline{l_L} \gamma_\mu,  \ee
where $\lambda=-\sqrt{2} G_F V_{ud}^*$. Here we are interested in neutrino production through the DU processes. The relevant non-time-ordered 2-point functions $\Sigma^>(x_1,x_2)$ and $\Sigma^<(x_1,x_2)$ are represented by the diagrams in Fig.~\ref{diagramF} and the corresponding analytic expressions can be written using the Kobes-Semenoff rules~\cite{KS,KS2} (except that the 2-point functions $S^>$ and $S^<$  should be those with generic  $\vec\Omega$ and chemical potentials derived in Sec.~\ref{Thermal propagator f}). Inserting $\Sigma^>(x_1,x_2)$ and $\Sigma^<(x_1,x_2)$ associated with the diagrams in Fig.~\ref{diagramF} in~(\ref{probF}) and~(\ref{probAF}), one obtains the neutrino production rate through the DU processes.

One can take into account in-medium effects through mean-field methods~\cite{Liu:2001iz,Fu:2008zzg}. The result is the following. First, the masses of $n$ and $p$ have to be substituted with effective masses. Second, the energies of $n$ and $p$ should be shifted by appropriate potentials, $U_n$ and $U_p$, respectively, although only in the Fermi-Dirac distributions and in the  delta functions that implement energy conservation. Unfortunately, this picture is fully  understood only for non-rotating plasmas and it is not clear how the angular velocity can change the numerical values of the effective masses.
 Therefore, in the rest of this section, an analytic understanding of  the Direct Urca processes is presented without performing a numerical analysis.

A first thing to notice is that the discussion of Sec.~\ref{Fermi momentum and Fermi surface} is valid even taking into account in-medium effects through mean-field methods. This is because one can interpret $M_i$ and $\mu_i$ there as the effective mass and the effective chemical potential of the $i$-th fermion species. The term effective chemical potential refers here to the difference between the actual chemical potential and the corresponding mean-field potential (e.g.~$U_n$ and $U_p$ for neutrons and protons, respectively).

An interesting property of the neutrino production rate through the DU processes is that it grows indefinitely as $\Omega\to 1/R$. The reason is the following. Let us consider the sum over the angular-momentum quantum numbers
of the fermions that take part in the processes
and focus on the behavior of the terms in the sum for large angular-momentum quantum numbers. Those are integrals involving the product of three  Fermi-Dirac distributions of fermions at thermal equilibrium.  
Each Fermi-Dirac distribution has the form
\be f_F(\omega_i - \Omega (m_i+1/2)-\mu_i), \ee
if the corresponding $i$-th fermion is in the initial state
and 
\be f_F(\mu_f-\omega_f + \Omega (m_f+1/2)), \ee
if the corresponding $f$-th fermion is in the final state. Here, $\{\omega_i, m_i, \mu_i\}$ and $\{\omega_f, m_f, \mu_f\}$ are the energies (computed with effective masses), angular-momentum quantum numbers and effective chemical potentials of the fermions at thermal equilibrium in the initial and final states respectively.
Now, these Fermi-Dirac distributions help the convergence of the sum over the $m_i$ and  $m_f$ only if
\be \Omega <  \lim_{m_i\to +\infty}\frac1{m_i}\min_{n_i\,p_i} ( \omega_{m_i,n_i}(p_i)) = \lim_{m_i\to +\infty} \frac{j_{m_i,1}}{m_i R} = \frac1{R}.   \label{ConvCond3} \ee 
Here $n_{i,f}$ is the other quantum number  of the fermion associated with the energy discretization at finite volume and  $p_{i,f}$ is its momentum along $\vec\Omega$.
A detailed inspection of the neutrino production rate, using e.g.~the explicit expression of the  $\mathcal{U}_q$ and $\mathcal{V}_q$ given by~\cite{Ambrus:2015lfr}, shows that  the large-$m_{i,f}$ term in the sum over the $m_{i,f}$ has the form of sums over the  $n_{i,f}$ and the $p_{i,f}$ of the above-mentioned Fermi-Dirac distributions times quantities that do not go to zero as the energies and the linear  momenta of the involved fermions go to infinity. So the sums over the energies and the linear  momenta do not converge as $\Omega\to 1/R$.
In other words, the neutrino production rate through the DU processes grows indefinitely as $\Omega\to 1/R$. The conclusion is that rotation can increase the neutrino production rate due to the DU processes.

Eventually, one is interested in taking the $L\to\infty$ and $R\to\infty$ limit, where the full space is recovered, to remove any dependence on the shape of the finite-volume region. No infrared divergences are present in this limit in the neutrino production rate  per unit of volume, $\pi R^2 L$, due to the DU processes. This can be explicitly checked by looking in detail at the neutrino production rate in question and using the results in the appendix
for the integral of products of Bessel functions\footnote{Moreover, it has been checked that the neutrino production rate per unit of volume in question reduces for $v\to 0$ to the known expression in the literature~\cite{Lattimer:1991ib}.}.

Moreover, one can show that for typical temperatures of neutron stars, where the energy and momentum of the (anti)neutrino is negligible, the condition\footnote{Here $\mu_n$, $\mu_p$ and $\mu_l$ are the chemical potentials of $n$, $p$ and $l$, respectively,  while the effective chemical potentials are obtained by subtracting the corresponding mean-field potentials. This notation will allow us to show that the condition for beta equilibrium depends neither on the angular velocity nor on the mean-field potentials.} $\mu_n=\mu_p+\mu_l$  ensures that the average rates of the  processes in~(\ref{DirectUrca}) are equal even for rotating neutron stars. Indeed, if the first process in~(\ref{DirectUrca}) had a larger (smaller) rate one would quickly have $\mu_n<(>)\mu_p+\mu_l$, which would block (favour) that process and favour (block) the other one until $\mu_n=\mu_p+\mu_l$ is restored.  To understand the latter statement, note  that in the first process  in~(\ref{DirectUrca}) the corresponding rate features \be f_F(\omega_n+U_n-vy_n-\mu_n)f_F(\mu_p-\omega_p-U_p+v y_p)f_F(\mu_e-\omega_e+v y_e),\label{fFinndec}\ee where $\{\omega_n, y_n\}$, $\{\omega_p, y_p\}$ and $\{\omega_e, y_e\}$ are the values of $\{\omega, y\}$ for $n$, $p,$ and $e$, respectively. Also, note that energy and momentum conservation (neglecting the neutrino energy and momentum) implies respectively 
\bea \omega_n+U_n &=& \omega_p+U_p + \omega_e, \\
y_n &=& y_p + y_e \eea 
and so
$$\omega_n+U_n-vy_n = \omega_p+U_p-vy_p + \omega_e-vy_e.$$ 
On the other hand, the combination of Fermi-Dirac distributions in~(\ref{fFinndec}) and the strong $n$, $p$ and $e$ degeneracy for typical temperatures of neutron stars tells us that the rate is strongly suppressed unless $\omega_n+U_n-vy_n\lesssim\mu_n$, $\omega_p+U_p-vy_p\gtrsim\mu_p$ and $\omega_e-vy_e\gtrsim\mu_e$. Analogously, one finds that the rate of the second process in~(\ref{DirectUrca}) is strongly suppressed unless $\omega_n+U_n-vy_n\gtrsim\mu_n$, $\omega_p+U_p-vy_p\lesssim\mu_p$ and $\omega_e-vy_e\lesssim\mu_e$.  So $\mu_n=\mu_p+\mu_l$ is, for any $v\in[0,1)$, the condition for beta equilibrium, with no dependence on the angular velocity.  

\section{Summary and conclusions}\label{Conclusions}

Let us conclude by providing a summary of the main original results obtained.
\begin{itemize}
\item In Sec.~\ref{fermion Free fields} the analysis of fermion TFTs for a generic equilibrium density matrix started with the simplest case of free fermions (or quasi-free fermions, when in-medium effects are taken into account with effective masses and effective chemical potentials). Sec.~\ref{fermion Free fields} included the most general spin-1/2 particle content (covering both Dirac and Majorana fermions), featuring generic masses (covering both Dirac and Majorana masses), chemical potentials and thermal vorticity (corresponding to the average angular momentum).  In order to describe both Dirac and Majorana fermions in the most convenient way both Dirac spinors and Weyl spinors were used.

In Sec.~\ref{Computing ensemble averages f} the averages of the product of two annihilation and creation operators were derived in a closed form, which allowed us to compute the averages of $H$, $\vec J$ and $Q^a$. An important finding is that the convergence of the averages requires $\Omega<1/R$. The large-volume limit can be taken by keeping $v\equiv R\Omega\in[0,1)$ fixed, but that convergence requirement implies that the averages $\langle \rho_E\rangle$, $\langle\mathcal{J}_z\rangle$ and $\langle \rho_a\rangle$ in~(\ref{rhoEpf}),~(\ref{CallJzf}) and~(\ref{rhoaf}) increase indefinitely as 
$v$ approaches $1$. Thus, by varying $v\in[0,1)$ at fixed temperature and chemical potentials one can obtain all values of the average angular momentum, then the average energy and charges are predicted. Special attention was devoted to the case of a strongly degenerate Dirac fermion, which is relevant for neutron stars.

In Sec.~\ref{Thermal propagator f} the thermal propagator and the ``non-time-ordered" 2-point functions of a fermion field in a generic irreducible representation  with arbitrary chemical potentials and thermal vorticity were obtained. Again both Dirac and Majorana fermions were covered. This was done by exploiting the averages of the product of two annihilation and creation operators previously calculated. Like in non-statistical quantum field theory, the thermal propagator is an important ingredient to perform perturbation theory in an interacting perturbative theory. 
\item The description of a generically interacting theory was then provided in Sec.~\ref{Fermion path integral} by deriving path-integral expressions for the partition function  and the thermal Green's functions for the most general fermion-scalar theory. A formalism which includes the real- and imaginary-time formalism was adopted, by generalizing existing methods to include the average angular momentum and providing formul\ae~that are applicable in perturbation theory (and beyond if the chemical potentials are not present). 

\item Sec.~\ref{Some applications} provided some applications of the previously obtained results, paying special attention to cases of relevance for neutron stars.

 One application was the determination (given in Sec.~\ref{Fermi momentum and Fermi surface}) of the Fermi surface for general angular velocity as well as the corresponding momenta. Interestingly, it was found that the angular velocity favors the existence of a Fermi surface and the maximum momentum on this surface grows indefinitely  as $v\to 1$. Unlike for $v\neq 0$, the modulus of the linear momentum on the Fermi surface can acquire several values.

 Sec.~\ref{Weakly-coupled fermion production} then presented general expressions to compute the production rates of weakly-coupled fermions from a rotating plasma featuring arbitrary values of chemical potentials and temperature. Both Dirac and Majorana fermions were covered. The latter can, for example, describe sterile neutrinos in a type-I  see-saw model.  
 
 These expressions were then applied to the DU processes for neutrino emission in rotating neutron stars in Sec.~\ref{Direct URCA processes in rotating neutron stars}. It was analytically shown that the neutrino production rate through the DU processes grows indefinitely as $\Omega\to 1/R$. Furthermore, it was analytically shown that such rate does not suffer from infrared divergences in the large-volume limit, $L\to \infty$ and $R\to\infty$ with $v\equiv \Omega R$ fixed. This is useful because it allows us to obtain formul\ae~that can be applied (through integration) to rotating plasmas of any shape. To achieve this result it was used a technique to represent the integral of the product of a general number of cylindrical Bessel functions, which is presented in the appendix; this representation could be useful in the future, for example, to numerically compute particle interaction rates in the presence of rotation.  Finally, it was found that the condition for beta equilibrium does not depend on the angular velocity of the rotating plasma.

 \end{itemize}

 As an outlook, one could apply the results of this work to compute, for example, the average  energy, number and angular momentum as well as the production rate of several types of particles from other rotating compact objects, including astrophysical or primordial rotating black holes.
 
\vspace{0.6cm}
 
 \subsection*{Acknowledgments}
I thank Francesco Tombesi for valuable discussions on rotating plasmas around black holes.

  \appendix

 \section{Integral of products of Bessel functions}\label{Integral of products of Bessel functions}
  
  In this appendix it is discussed a strategy to  compute 
  integrals of the form
  \be I_{q_0,q_1,q_2, ... , q_N}\equiv \int_0^{\infty} dr r \prod_{j=0}^NJ_{m_j}(\alpha_j r), \ee
  which appear in the rates when a non-vanishing average angular momentum is present. Here, $J_m(z)$ is the cylindrical Bessel function of argument $z$ and order $m$ and the $q_j$, with $j=0,1,2, ... , N$, correspond to the pairs $\{\alpha_j, m_j\}$, where the $\alpha_j$ are non-negative real numbers and the $m_j$ are integers. 
  
  Let us start from the generating function of the cylindrical Bessel functions~\cite{Watson}:
  \be e^{z(t-1/t)/2}= \sum_{m=-\infty}^{+\infty} t^m J_m(z).\ee
  Setting $t=i \exp(i\theta)$, 
  \be e^{iz\cos\theta} =  \sum_{m=-\infty}^{+\infty}i^m e^{im\theta} J_m(z). \label{GenStep}\ee 
  Now, one can interpret $\theta$ as the angle between two bi-dimensional vectors, $\vec\alpha$ and $\vec r$ (which in polar coordinates read $\vec\alpha=\{\alpha,\theta_\alpha\}$ and $\vec r=\{r,\theta_r\}$). Also, one can set $z=\alpha r$. So~(\ref{GenStep}) gives
  \be e^{i\vec\alpha\cdot\vec r} = \sum_{m=-\infty}^{+\infty}i^m e^{im\theta_r}e^{-im\theta_\alpha} J_m(\alpha r).\ee
  Multiplying both sides of the expression above by $\exp(im'\theta_\alpha)$ and integrating over $\theta_\alpha$ one obtains
  \be J_m(\alpha r) e^{i m \theta_r} = \frac{i^{-m}}{2\pi}\int d\theta_\alpha e^{i\vec\alpha\cdot\vec r}e^{im\theta_\alpha},\ee 
  where henceforth the integral over any angular variable is on the interval $[0,2\pi)$. 
  This relation implies 
  \be I_{q_0,q_1,q_2, ... , q_N} \int  d\theta_r  e^{i \sum_{j=0}^Nm_j\theta_r} = \frac{i^{-\sum_{j=0}^Nm_j}}{(2\pi)^{N+1}}\int d^2 r \prod_{j=0}^N d\theta_{\alpha_j} e^{i\vec\alpha_j\cdot \vec r}  e^{im_j \theta_{\alpha_j}},\ee 
  where $d^2 r \equiv r dr d\theta_r$. Performing the integral over $\theta_r$ on the left-hand side and over $\vec r$ on the right-hand side leads to
    \be I_{q_0,q_1,q_2, ... , q_N} \delta_{0,\sum_{j=0}^Nm_j}=\frac{i^{-\sum_{j=0}^Nm_j}}{(2\pi)^{N}}\int \delta\left(\sum_{j=0}^N\vec\alpha_j\right) \prod_{j=0}^N d\theta_{\alpha_j}  e^{im_j \theta_{\alpha_j}}. \ee
  
  This relation allows us to compute $I_{q_0,q_1,q_2, ... , q_N} $ when 
  \be \sum_{j=0}^Nm_j = 0. \label{AngCon}\ee
  In our case
   this condition is not restrictive because it is always satisfied thanks to angular momentum conservation along $\vec\Omega$.
   Using~(\ref{AngCon}) leads to
  \be I_{q_0,q_1,q_2, ... , q_N} =\frac{1}{(2\pi)^{N}}\int \delta\left(\sum_{j=0}^N\vec\alpha_j\right) \prod_{j=0}^N d\theta_{\alpha_j}  e^{im_j \theta_{\alpha_j}}. \ee
This results shows that, as long as~(\ref{AngCon}) holds,  
$I_{q_0,q_1,q_2, ... , q_N}$ vanishes unless the $\alpha_j$, with $j=0, 1, 2, ..., N$, can form the sides of a polygon.  

 One can easily perform one of the angular integration by exploiting the rotational invariance of the Dirac delta function and the measures $d\theta_{\alpha_j}$. 
One defines $\zeta_j\equiv \theta_{\alpha_j} - \theta_{\alpha_0}$  for $j\neq 0$, to obtain
  \be I_{q_0,q_1,q_2, ... , q_N} =\frac{1}{(2\pi)^{N-1}}\int \delta\left(\sum_{j=0}^N\vec\alpha_j\right) \prod_{j=1}^N d\zeta_j  e^{im_j \zeta_j}, \label{GenI}\ee
  where Condition~(\ref{AngCon}) was used again. 
  
  Note that for $N=1$ the result in~(\ref{GenI}) reduces to 
  \be I_{q_0,q_1} = (-1)^{m_1}\frac{\delta(\alpha_1-\alpha_0)}{\alpha_0}, \ee 
  which agrees with the closure relation of cylindrical Bessel functions.
 The case $N=2$ was discussed in~\cite{JacksonMaximon}.  Higher values of $N$ can be addressed as follows. First, define the bi-dimensional vector
 \be \vec p \equiv \sum_{j=0}^{N-2} \vec\alpha_j. \ee 
 In the rotated reference frame, where the $\vec\alpha_j$ have angular polar coordinates $\zeta_j$, in general $\vec p$ has a non vanishing angular polar coordinate, which will be called $\zeta_p$. Note that $\zeta_p$ generically depends on the $\zeta_j$ for $j=1, ..., N-2$. 
One can now write
    \be I_{q_0,q_1,q_2, ... , q_N} =\frac{1}{(2\pi)^{N-1}}\int \left(\prod_{j=1}^{N-2} d\zeta_j  e^{im_j \zeta_j}\right)\int  d\zeta_{N-1}  d\zeta_N e^{im_{N-1} \zeta_{N-1}+im_{N} \zeta_{N}}\delta\left(\vec p+\vec\alpha_{N-1}+\vec\alpha_{N}\right). \label{GenI2}\ee
    Second, define the new angles $\zeta'_{N-1}\equiv \zeta_{N-1}-\zeta_p$ and $\zeta'_{N}\equiv \zeta_{N}-\zeta_p$ and exploit again the rotational invariance of the Dirac delta function and the measures $d\zeta_{N-1}$  and $d\zeta_N$ to obtain
  \be   I_{q_0,q_1,q_2, ... , q_N} =\frac{1}{(2\pi)^{N-1}}\int \left(\prod_{j=1}^{N-2} d\zeta_j  e^{im_j \zeta_j}\right)e^{i(m_{N-1}+m_N) \zeta_p} \int  d\zeta'_{N-1}  d\zeta'_N e^{im_{N-1} \zeta'_{N-1}+im_{N} \zeta'_{N}}\delta\left(\vec p+\vec\alpha_{N-1}+\vec\alpha_{N}\right). \nonumber  \ee
  This result can be rewritten as
\be I_{q_0,q_1,q_2, ... , q_N} =\frac{1}{(2\pi)^{N-2}}\int \left(\prod_{j=1}^{N-2} d\zeta_j  e^{im_j \zeta_j}\right) e^{i(m_{N-1}+m_N) \zeta_p} I_{q_p, q_{N-1}, q_N}, \label{GenI3}\ee
where $q_p$ corresponds to the pair $\{|\vec p|, -m_{N-1}-m_N\}$. So $I_{q_0,q_1,q_2, ... , q_N}$ for any $N$ can be computed by integrating over $N-2$ angular variables an expression that can be  determined with the known formula for $I_{q_0,q_1,q_2, ... , q_N}$ with $N=2$. 

\vspace{1cm}

 \footnotesize
\begin{multicols}{2}

\end{multicols}

  \end{document}